\documentclass[lettersize,journal]{IEEEtran}
\usepackage{amsmath,amsfonts}
\usepackage{algorithm}
\usepackage{array}
\usepackage[caption=false,font=normalsize,labelfont=sf,textfont=sf]{subfig}
\usepackage{textcomp}
\usepackage{stfloats}
\usepackage{url}
\usepackage{verbatim}
\usepackage{graphicx}
\usepackage{textcomp}
\usepackage{colortbl}
\usepackage{multirow}
\usepackage{xcolor}

\usepackage{array,booktabs,makecell}
\usepackage{cite}
\usepackage{algorithm}       
\usepackage{algpseudocode} 
\begin{document}

\title{Sub-array Selection Optimization for Joint Self-Interference and Multi-User Interference Suppression in FD mMIMO}

\author{{Yuanzhe Gong,~\IEEEmembership{Member,~IEEE}, Yuanxing Zhang, Tho Le-Ngoc,~
\IEEEmembership{Life~Fellow,~IEEE}
}

\thanks{ Y. Gong, Y. Zhang, and T. Le-Ngoc are with the Department of Electrical and Computer Engineering, McGill University, QC, Montreal, Canada (Emails:
 \{yuanzhe.gong, yuanxing.zhang2\}@mail.mcgill.ca, tho.le-ngoc@mcgill.ca).
 }}

\markboth{Journal of \LaTeX\ Class Files,~Vol.~a, No.~a, August~2025}%
{Shell \MakeLowercase{\textit{et al.}}: A Sample Article Using IEEEtran.cls for IEEE Journals}


\maketitle

\begin{abstract}
This paper proposes a beamforming optimization scheme with joint antenna sub-array selection (SAS) and angular perturbation-based nulling (APN) for full-duplex (FD) massive multiple-input multiple-output (mMIMO) systems, to simultaneously suppress self-interference (SI) and multi-user interference (MUI). A comprehensive over-the-air SI channel measurement campaign, conducted with an $8\times8$ Tx-$8\times8$ Rx FD array prototype, reveals significant variations across sub-arrays at different spatial locations, as well as reconfigurable characteristics of the SI channel under diverse Tx and Rx sub-array configurations. To exploit the selective SI channels, a particle swarm optimization (PSO)-based algorithm is developed to jointly determine optimal sub-array indices and perturbed steering angles, thereby effectively nullifying potential interference. Selecting sub-arrays with inherently lower SI channels notably enhances the beam-level isolation, while the added selection flexibility among comparable SI channels ensures more uniform SI suppression across diverse DL/UL locations and significantly improves worst-case isolation. Experimental evaluation based on the measured SI channel demonstrates that the proposed SAS technique achieves residual Tx-Rx beam-level SI suppression improvements of 29.2 dB and 26.6 dB for the sample $1\times2$ and $1\times4$ sub-arrays, respectively. A worst-case improvement greater than 30.7 dB is observed, representing a substantial isolation gain under the SI bottleneck scenario. Overall, the joint SAS and APN optimization scheme achieves average beam-level isolation of 85.2 dB and 83.3 dB with the $1\times2$ and $1\times4$ sub-arrays, respectively. With the application of a baseband precoder, all tested sub-array configurations achieve average MUI suppression better than -181.3 dB. These results confirm the potential of the proposed optimization algorithm to successfully reduce interference to the noise floor, thereby guaranteeing reliable FD mMIMO operation.

\end{abstract}


\section{Introduction}
\IEEEPARstart{M}{assive} multiple-input multiple-output (mMIMO) was proposed over a decade ago as a transformative advancement in wireless communications \cite{rusek2012scaling, marzetta2010noncooperative, bjornson2019massive}.
By equipping base stations (BSs) with large-scale antenna arrays, multi-user (MU) mMIMO substantially enhances both spectral efficiency and network capacity, by concurrently supporting an increased number of data streams in different spatial directions \cite{agiwal2016next, shafi20175g, busarimillimeter, marzetta2010noncooperative, rusek2012scaling}. 
Through highly focused and strategically controllable beams, spatial separation among users is enhanced, thereby improving signal quality and enabling more effective interference management in scenarios with multiple active data streams \cite{YG_OJVT_2, bjornson2017massive}. 
In conventional mMIMO systems, downlink (DL) and uplink (UL) transmissions are separated either by frequency, as in frequency-division duplexing (FDD), or by time, as in time-division duplexing (TDD). 
By contrast, full-duplex (FD) communication enables simultaneous transmission and reception of signals within the same frequency band, potentially doubling spectral efficiency compared to conventional half-duplex (HD) systems \cite{zhang2016full, sabharwal2014band}. 
By supporting concurrent UL and DL services, the integration of FD with mMIMO systems can provide additional gains, including enhanced energy and spectral efficiency, higher network capacity, and reduced latency \cite{sabharwal2014band, bharadia2013full, duarte2013design}.

Despite the advantages of FD communication, a major obstacle to its widespread adoption is the strong self-interference (SI) generated by its own concurrent transmitter (Tx) \cite{zhang2016full, kim2015survey, kolodziej2019band}. 
The SI is often much stronger than the intended signal, which can overload the receiver (Rx) and severely degrade the received signal-to-interference-plus-noise ratio (SINR), ultimately limiting channel throughput \cite{chen2017rf}.
To achieve detection performance comparable to HD systems, the residual SI must ideally be suppressed below the thermal noise floor. For instance, assuming a transmit power of \( P_T = 30\, \text{dBm} \), a noise power spectrum density of \(-174\, \text{dBm/Hz}\), and an operating bandwidth of 20 MHz, an SI suppression $131 \, \text{dB}$ is needed  \cite{Report_5G_Macro_PL_Rel_17, Report_5G_UMi_UMa_Rel17}.
Achieving such a substantial level of SI suppression presents a significant technical challenge in FD mMIMO system design.

Tx-Rx isolation, which separates transmitted and received signals, is a critical first step in mitigating Tx-Rx mutual coupling (MC) and preventing the overloading of front-end components such as analog-to-digital converters (ADCs) and low-noise amplifiers (LNAs). 
RF isolation can be enhanced either by increasing the physical separation between the Tx and local Rx arrays, thereby introducing higher path loss for the SI signal \cite{duarte2013design, slingsby1995antenna}, or by employing cross-polarized Tx-Rx designs that maximize the immunity of local Rx antennas to leakage coupling waves \cite{debaillie2014analog, nawaz2017dual, ha2017monostatic, chen2018self, wu2020compact, YG_VTC_2}. 
Various coupling-suppression structures, such as frequency-selective surfaces \cite{ghosh2014dual, farahani2010mutual, ahn2001design} and metamaterial absorbers \cite{zhai2015enhanced, dadgarpour2016mutual, YG_ACCESS_2}, have also been developed to attenuate or dissipate residual coupling.

Moreover, large-scale arrays further enhance SI isolation by enabling narrow, steerable beams that minimize leakage power in unintended directions \cite{everett2014passive}. 
RF beamformer optimization can then refine UL and DL beam interactions by directing beams toward interference-nulling locations, thereby achieving additional SI suppression \cite{koc2022intelligent, YG_ACCESS_1, YG_ACCESS_2,gong2024nulling}. 
To reduce hardware complexity while effectively leveraging large-scale arrays for FD mMIMO SI management, hybrid beamforming (HBF) has gained prominence for joint DL/UL beamformer design \cite{rihan2020taxonomy}. 
For instance, \cite{huberman2014mimo} proposes a sequential convex programming approach for separate and joint FD beamforming that maximizes sum rates in both single-user (SU) and MU-MIMO FD systems. 
Similarly, \cite{satyanarayana2018hybrid} develops an iterative FD HBF scheme that jointly designs transmit/receive RF beamformers, precoders, and combiners, achieving up to 30 dB of SI suppression.
HBF optimization schemes are developed in \cite{koc2021full, koc2022intelligent, YG_ACCESS_2} that exploit angle-of-departure and angle-of-arrival information for both near-field and far-field SI mitigation.

Although existing multi-stage SI suppression techniques in FD mMIMO systems \cite{koc2022intelligent, huberman2014mimo, zhang2019precoding, luo2021robust, koc2021full, satyanarayana2019multi, darabi2021transceiver} have demonstrated effectiveness in mitigating Tx-Rx SI, achieving isolation levels of 78 to 81 dB \cite{koc2021full,koc2022intelligent}, their performance has primarily been evaluated using theoretical SI channel models and simulations, generally assuming fixed antenna configurations and static SI-channel conditions.
For example, \cite{koc2022intelligent, huberman2014mimo, zhang2019precoding, luo2021robust, koc2021full} base their assumptions and analyses on simulated near-field SI channels over line-of-sight (LoS) paths and far-field SI channels via reflected non-line-of-sight (NLoS) paths.
However, owing to the close proximity between the Tx and Rx arrays, the SI channel is governed by multiple leakage mechanisms, including near-field MC, surface-wave leakage, and spatial leakage. Furthermore, practical RF-isolation structures employed in the antenna and RF design also shape the SI response. As a result, theoretical SI models alone are generally insufficient to accurately represent realistic SI conditions or to fully capture the SI environment targeted by the proposed design.
Therefore, an evaluation of beamforming-based SI suppression that incorporates realistic electromagnetic interactions within and between arrays is essential for validating FD mMIMO implementations.
Furthermore, steering-angle-based isolation approaches often suppress SI at the cost of desired signal quality, since main lobes need to be steered away from target user locations to align with interference-nulling directions. 
Additionally, the effectiveness of such methods is also constrained by the strength of the Tx-Rx beam interaction, making beam-level SI suppression highly dependent on the distribution of DL and UL users.
More than 20 dB of beam-level isolation variation can be observed across different DL/UL user locations \cite{YG_ACCESS_1,mahmood2024achieving,YG_OJVT_1}.

Recent research in multiple-antenna systems has highlighted the potential of selective and reconfigurable signal propagation channels, which offer greater flexibility in selecting optimal antennas to mitigate SI based on user locations, while simultaneously maximizing the desired signal quality \cite{yang2015efficient,fidan2018performance,zhu2022antenna,wilson2017antenna,jang2016antenna,mahmood2024achieving,ding2024movable,ding2024movable2,gao2015massive}.
A joint relay and transmit/receive antenna mode selection (RAMS) scheme has been proposed in \cite{yang2015efficient} for FD relay networks, where each relay adaptively selects its transmit and receive antennas according to instantaneous channel conditions, and the optimal relay configuration is chosen to maximize the end-to-end SINR. 
Analytical and simulation results show that the scheme effectively eliminates the SI-induced error floor while providing additional spatial diversity.
In \cite{wilson2017antenna}, antenna selection algorithms for FD communications were developed based on magnitude, orthogonality, and determinant criteria. These algorithms achieve near-optimal performance with low complexity, favouring single-antenna transmission at low signal-to-noise ratio (SNR) and balanced Tx-Rx selection at high SNR. 
Later, \cite{fidan2018performance} proposed a novel transceiver antenna selection strategy for FD MIMO systems, where transmit/receive antenna pairs at terminal sources are selected based on the maximum and second maximum SINR. 
This strategy achieves up to 25 dB gain over non-selection cases and provides diversity orders dependent on SI, power allocation, and relay location.
For FD distributed mMIMO systems, \cite{zhu2022antenna} formulated antenna selection as a spectrum-efficiency maximization problem and solved it using an elite preservation genetic algorithm (EPGA), which achieves near-optimal performance and significantly outperforms random assignment. Similarly, \cite{jang2016antenna} investigated antenna selection methods for bidirectional FD MIMO systems, where a proposed low-complexity algorithm attains near-optimal average sum-rate performance, yielding up to 15\% gain compared with conventional schemes. 
Recently, additional degrees of freedom in jointly optimizing the desired and interference channels have been achieved through a movable-antenna FD BS architecture combined with an alternating optimization framework that jointly optimizes antenna positions, beamformers, and UL powers to maximize secrecy rates \cite{ding2024movable,ding2024movable2}.

However, most existing studies \cite{yang2015efficient,fidan2018performance,wilson2017antenna,jang2016antenna,ding2024movable,ding2024movable2} on antenna selection or reconfigurable-position antennas for FD communications have primarily focused on systems with a limited number of antennas, often restricted to single-antenna DL/UL user scenarios.
For mMIMO, antenna selection over large-scale arrays can substantially enhance system performance. Measurements in \cite{gao2015massive} on 128-element HD arrays show that even simple power-based selection achieves performance close to convex optimization benchmarks, while simultaneously reducing the number of required RF chains and overall system complexity.
Therefore, there is a pressing need to extend antenna and sub-array selection techniques to FD communication with large-scale antenna arrays, where strategic beamforming optimization can be leveraged to further improve desired signal quality and effectively manage interference. 
Moreover, the lack of evaluation based on experimentally measured SI channels in \cite{yang2015efficient,fidan2018performance,zhu2022antenna,wilson2017antenna,jang2016antenna,ding2024movable,ding2024movable2} limits the practical validation of channel-reconfigurable or selection-based algorithms. In particular, reconfigurable SI-channel selection in large-scale FD arrays has not been systematically studied under realistic SI conditions, especially when jointly considered with RF beamforming design.
Although \cite{mahmood2024achieving} proposed a hybrid beamforming architecture combining non-orthogonal beamforming and sub-array selection for FD communication, achieving up to $-78$ dB SI suppression, the study was restricted to single-DL and single-UL scenarios and therefore did not consider multi-user interference (MUI), which is another major limiting factor and can be effectively mitigated through proper RF beamformer optimization.
Therefore, further investigation is still required to characterize reconfigurable SI-channel distributions, clarify the mechanism by which sub-array selection improves RF isolation, and evaluate the SI-suppression performance of different sub-array lengths together with their impact on MUI mitigation in FD MU-mMIMO systems. 
In particular, SI-channel magnitude variation and spatial diversity across different Tx and Rx sub-array configurations have not yet been systematically characterized. Such analysis is essential for understanding the mechanism and practical effectiveness of sub-array selection for SI suppression.

Following the RF/analog SI-isolation stage, the residual SI can be further mitigated using digital baseband self-interference cancellation (SIC) techniques \cite{luo2022design,dastjerdi201928,liu2023joint}. For instance, a 16-tap baseband canceller achieves approximately 50 dB of additional suppression after RF SIC in \cite{luo2022design}. 
Likewise, \cite{dastjerdi201928} reports up to 45 dB average digital SIC over a 20 MHz bandwidth using a delay-based finite impulse response (FIR) filtering canceller in a 2$\times$2 FD MIMO setup. 
In \cite{liu2023joint}, a penalty-based iterative algorithm is developed for an FD monostatic integrated sensing and communication (ISAC) system. By jointly optimizing the transceiver beamformers, UL precoder, and DL combiner, the proposed design achieves up to 60 dB of digital-domain SIC, while also improving the average sum-rate and estimation accuracy.
Nevertheless, RF isolation remains the primary line of defence against SI. To bring the residual SI close to the noise floor, an overall isolation of more than 80 dB is typically required prior to the above digital cancellation.

\subsection{Contributions}
Building on the aforementioned observations, this study aims to develop a joint sub-array selection (SAS) and angular perturbation-based nulling (APN) beamforming optimization scheme to effectively mitigate both SI and MUI in FD mMIMO systems.
Furthermore, a comprehensive measurement-based investigation of the SI channel between Tx and Rx large-scale antenna array prototypes is conducted, which is further leveraged to explore the degrees of freedom and potential of reconfigurable SI channel selection for enhanced SI suppression.
The main contributions of this work are summarized as follows:
\begin{itemize}
  \item \textbf{Comprehensive $8\times8$Tx-$8\times8$Rx Measurement-Based SI Channel:} 
  A complete experimental measurement of the SI channel between two $8\times8$ Tx and $8\times8$ Rx uniform rectangular arrays (URAs), incorporating our previously developed metamaterial-based patch antenna elements \cite{YG_GC_22,YG_ACCESS_1}, is conducted and presented. The realistic element-level SI-channel variations across different relative spatial locations, the spatial correlation among candidate SI sub-channels, and the distribution of sub-array SI coupling magnitudes within the Tx and Rx large-scale arrays are systematically analyzed. These results reveal the reconfigurable characteristics of the SI channels under different Tx and Rx sub-array configurations, elucidate the available design freedom and the distinct mechanisms associated with different sub-array structures, and identify favourable SI-channel conditions for FD-mMIMO operation.

    \item \textbf{Sub-array Selection Optimization Scheme for SI Suppression Improvement:} An antenna sub-array selection optimization approach is developed to leverage the opportunities offered by large-scale arrays in FD communications. Experimental evaluation based on the measured SI channel demonstrates a significant enhancement in SI suppression when employing the SAS technique, achieving residual Tx-Rx beam-level SI suppression improvements of $29.2$ dB and $26.6$ dB for the sample $1\times2$ and $1\times4$ sub-arrays, respectively. 
    A worst-case improvement exceeding $30.7$ dB ($1\times2$ sub-array) and $28.3$ dB ($1\times4$ sub-array) is achieved, indicating a significant isolation gain in the SI bottleneck scenario.
    Selecting sub-arrays associated with inherently lower SI channels results in a pronounced improvement in the average beam-level isolation. 
    Moreover, the increased selection flexibility among comparably performing channels leads to more uniform SI suppression across various DL/UL locations and substantially improves the worst-case isolation.

    \item \textbf{Joint Reconfigurable SI Channel Selection and Hybrid Beamformer Optimization for FD mMIMO:}
    A joint angular perturbation-based null-space projection and sub-array selection beamforming optimization scheme is proposed to maximize the spatial isolation capability of RF beamformer design by selecting the optimal sub-array SI channels, and aligning and projecting nulls toward potential interference signals. 
    The resulting design problem is mixed discrete-continuous, and a particle swarm optimization (PSO)-based approach is adopted for efficient solution. Because the optimization involves only the sub-array indices and their associated angular perturbations, the problem is low-dimensional, with the number of variables depending primarily on the number of users and candidate sub-arrays rather than on the total number of beamforming coefficients. Illustrative results demonstrate that the scheme achieves average beam-level SI suppression of $85.2$ dB and $83.3$ dB with the sample $1\times2$ and $1\times4$ sub-arrays, respectively, while maintaining Tx-Rx beam-level SI below $-78.4$ dB and $-73.4$ dB across all DL/UL user locations. Meanwhile, MUI levels remain below $-81.5$ dB for all tested sub-array configurations. Furthermore, after applying the baseband precoder, the system achieves a median MUI level better than $-185.7$ dB and an average MUI level better than $-181.3$ dB across the three sub-array schemes.

\end{itemize}

\subsection{Organization}
The remainder of the paper is organized as follows. Section~\ref{system_model} introduces the HBF system model for FD mMIMO systems, highlighting the challenges of SI and MUI. Section~\ref{Joint_SAS_HBF} formulates the optimization problem that combines sub-array selection, joint DL/UL steering angle perturbation, and null-space projection to enhance spatial isolation between Tx and Rx beams. 
A PSO algorithm is developed to solve the resulting non-convex problem. The residual MUI is further suppressed by a regularized zero-forcing (RZF) precoder. 
Then, Section~\ref{Meas_SI} presents the measured SI channels between $8\times8$ Tx and $8\times8$ Rx large-scale arrays and systematically analyzes the realistic SI channel variations across sub-arrays at different relative spatial locations. Section~\ref{Results} provides illustrative results based on these measured SI channels, where the residual SI and MUI performance across various DL/UL user combinations are investigated.
Finally, Section~\ref{Conclusions} concludes the paper and outlines potential directions for future research.

\subsection{Notation and Coordinates}
The following notations are used throughout this paper.
Square brackets $[\cdot]$ denote vectors and matrices, while parentheses $(\cdot)$ are used to indicate functional forms of variables, such as the interference channel matrix.
Bold uppercase letters represent matrices, and bold lowercase letters represent vectors.
$(\cdot)^T$, $(\cdot)^H$, $\lVert\cdot\rVert$, and $\mathrm{tr}(\cdot)$ denote the transpose, Hermitian transpose, 2-norm, and trace of a vector or matrix, respectively.
$\mathbf{I}_a$, $\mathbb{E}{(\cdot)}$, and $\operatorname{Re}{(\cdot)}$ denote the $a \times a$ identity matrix, the expectation operator, and the real part of a complex value/vector, respectively.
We use $x_k \sim \mathcal{CN}(0,\sigma^2)$ to indicate that $x_k$ is a circularly symmetric complex Gaussian random variable with zero mean and variance $\sigma^2$.
In the presentation, the antenna array is placed in the $yz$ plane and the main radiation direction is perpendicular to the array plane (along the $x$-axis) and is defined as $\theta=90^\circ$ and $\psi=90^\circ$.

\section{System Model}\label{system_model}
To address the challenges of high hardware cost and complexity, HBF has emerged as a promising solution for mMIMO systems \cite{sohrabi2016hybrid, el2014spatially}. Unlike fully digital beamforming (DBF), HBF divides the beamforming process into two stages: an analog stage operating at radio frequency (RF) and a digital stage operating at baseband. By interconnecting these stages through a reduced number of RF chains, HBF enables mMIMO systems to achieve near-DBF performance while remaining practical and cost-efficient for large-scale deployment \cite{ahmed2018survey}.

\subsection{Sub-Connected Hybrid Beamforming for FD MU-mMIMO}

\begin{figure}[!t]
\centerline{\includegraphics[width=\linewidth]{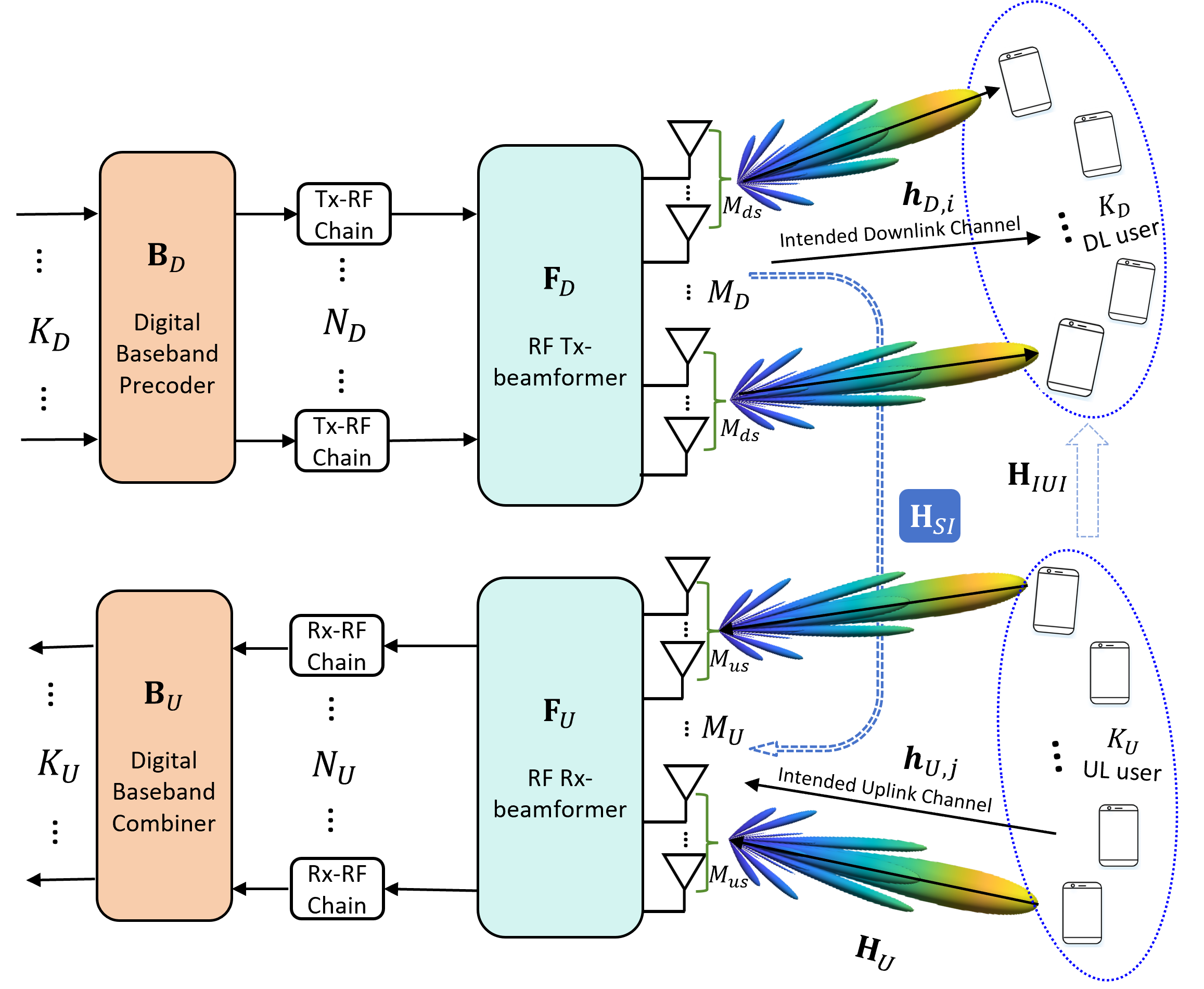}}
\caption{Sub-Connected HBF system model for FD MU-mMIMO.}
\label{FD_mMIMO_model}
\end{figure}

As illustrated in Fig.~\ref{FD_mMIMO_model}, an FD MU-mMIMO system is considered, where the BS employs a sub-connected hybrid beamforming (SC-HBF) architecture for both DL and UL.
The BS is equipped with $M_D$ transmit and $M_U$ receive antenna elements, which are equally partitioned into DL and UL sub-arrays consisting of $M_{ds}$ and $M_{us}$ elements, respectively.
This arrangement enables the BS to simultaneously serve $K_D$ single-antenna DL users and $K_U$ single-antenna UL users within the same time-frequency resources, subject to $K_D \leq \frac{M_D}{M_{ds}}$ and $K_U \leq \frac{M_U}{M_{us}}$.

On the DL, each user’s data symbol is first precoded by the baseband precoder $\mathbf{B}_D=[\mathbf{b}_{D,1},\dots,\mathbf{b}_{D,K_D}] \in \mathbb{C}^{N_D \times K_D}$, where $\mathbf{b}_{D,i} \in \mathbb{C}^{N_D \times 1}$ denotes the digital precoding vector for the $i^\text{th}$ DL stream. 
The precoded signals then pass through the RF beamformer $\mathbf{F}_D \in \mathbb{C}^{M_D \times N_D}$, where $N_D$ is the number of transmit RF chains and $K_D \leq N_D \ll M_D$.
This two-stage design reduces the number of RF chains and the dimension of the baseband precoder while maintaining sufficient spatial degrees of freedom for MU multiplexing.
Similarly, on the UL, the received signals are first combined by the RF beamformer $\mathbf{F}_U \in \mathbb{C}^{M_U \times N_U}$, followed by the baseband combiner $\mathbf{B}_U=[\mathbf{b}_{U,1},\dots,\mathbf{b}_{U,K_U}] \in \mathbb{C}^{N_U \times K_U}$, with $K_U \leq N_U \ll M_U$.

Each RF beamformer can be represented in block-diagonal form according to the angular directions $(\theta_{D_i}, \psi_{D_i})$ and $(\theta_{U_j}, \psi_{U_j})$ of the DL and UL sub-arrays,
 \begin{subequations}
 \begin{align}
 \mathbf{F}_D &= \begin{bmatrix}
\mathbf{f}_D(\theta_{D_1}, \psi_{D_1})   & \cdots & 0          \\
\vdots            & \ddots & \vdots     \\
 0                    & \cdots &\mathbf{f}_D(\theta_{D_{N_D}}, \psi_{D_{N_D}})
 \end{bmatrix}, \label{eq:FD} \\
 \mathbf{F}_U &= \begin{bmatrix}
 \mathbf{f}_U(\theta_{U_1}, \psi_{U_1})     & \cdots & 0          \\
 \vdots              & \ddots & \vdots     \\
 0                  & \cdots &\mathbf{f}_U(\theta_{U_{N_U}}, \psi_{U_{N_U}})
 \end{bmatrix}, \label{eq:FU}
 \end{align}
 \end{subequations}
where $\mathbf{f}_D(\theta_{D_i}, \psi_{D_i}) \in \mathbb{C}^{M_{ds} \times 1}$ and $\mathbf{f}_U(\theta_{U_j}, \psi_{U_j}) \in \mathbb{C}^{M_{us} \times 1}$ are the beamforming vectors for the $i^\text{th}$ and $j^\text{th}$ DL and UL sub-array, respectively. 
Both $\mathbf{F}_D$ and $\mathbf{F}_U$ are implemented using phase shifters, thereby imposing constant-modulus constraints on their entries, such that $|[\mathbf{f}_D]_m| = 1/\sqrt{M_{ds}}$ and $|[\mathbf{f}_U]_m| = 1/\sqrt{M_{us}}$.

The received signal at the $i^\text{th}$ DL user is expressed as
\begin{equation} 
\begin{split}\label{DL_signal}
r_{D,i} &= \underbrace{\mathbf{h}_{D,i}^H \mathbf{F}_{D} \mathbf{b}_{D,i} d_{D,i}}_{\text{Desired Signal}}  \hspace{1ex}
+ \hspace{0.5ex}\underbrace{\mathbf{h}_{\text{IUI},i}^H \mathbf{d}_{U}}_{\text{IUI from UL users}} \\
&\hspace{3ex}+ \underbrace{\sum_{\tilde{i} \neq i} \mathbf{h}_{D,{i}}^H \mathbf{F}_{D} \mathbf{b}_{D,\tilde{i}} d_{D,\tilde{i}}}_{\text{Multi-user Interference (MUI)}} 
\hspace{0.5ex} + \hspace{0.5ex} \underbrace{w_{D,i}}_{\text{Noise}},
\end{split}
\end{equation}
where $\mathbf{h}_{D,i} \in \mathbb{C}^{M_{D} \times 1}$ denotes the DL channel vector from the BS to the $i^\text{th}$ target user, 
$\mathbf{b}_{D,i}$ is the $i^\text{th}$ column of the baseband precoding matrix $\mathbf{B}_D$. 
The DL data vector is $\mathbf{d}_{D}=[d_{D,1},\dots,d_{D,K_D}]^T$ with normalized power constraint $\mathbb{E}\{\mathbf{d}_{D} \mathbf{d}_{D}^H\}=\mathbf{I}_{K_D}$.
$d_{D,i}$ is the data symbol sent to the $i^\text{th}$ user.
The second term models inter-user interference (IUI) from simultaneous UL transmissions, with $\mathbf{d}_{U}=[d_{U,1},\dots,d_{U,K_U}]^T \in \mathbb{C}^{K_U \times 1}$ and $\mathbb{E}\{\mathbf{d}_{U} \mathbf{d}_{U}^H\}=\mathbf{I}_{K_U}$, through the effective channel $\mathbf{h}_{\text{IUI},i} \in \mathbb{C}^{K_U \times 1}$. 
Due to the MU operation, the third term captures MUI from the transmit beam intended for user $\tilde{i}^\text{th}$ leaking into the $i^\text{th}$ DL user, where $\tilde{i} \neq i$. 
Finally, $w_{D,i} \sim \mathcal{CN}(0, \sigma^2)$ denotes the additive white Gaussian noise at the $i^\text{th}$ DL user.

Similarly, for the UL channel, the received signal at the BS from the $j^\text{th}$ UL user is given by
\begin{equation} 
\begin{split}\label{UL_signal}
r_{U,j} &= \underbrace{\mathbf{b}_{U,j}^H \mathbf{F}_{U}^H \mathbf{h}_{U,j} d_{U,j}}_{\text{Desired Signal}} + \underbrace{\sum_{\tilde{j} \neq j} \mathbf{b}_{U,j}^H \mathbf{F}_U^H \mathbf{h}_{U,\tilde{j}} d_{U,\tilde{j}}}_{\text{MUI}} \\
& + \underbrace{\mathbf{b}_{U,j}^H \mathbf{F}_{U}^H \mathbf{H}_{\text{SI}} \mathbf{F}_{D} \mathbf{B}_{D} \mathbf{d}_{D}}_{\text{Self-Interference (SI)}} + \underbrace{\mathbf{b}_{U,j}^H \mathbf{F}_U^H \mathbf{w}_{U}}_{\text{Noise After Combining}},
\end{split}
\end{equation}
where $\mathbf{h}_{U,j} \in \mathbb{C}^{M_{U} \times 1}$ is the UL channel vector from the $j^\text{th}$ user to the BS. 
$d_{U,j}$ is its transmitted data symbol, and $\mathbf{b}_{U,j}$ is the $j^\text{th}$ column of the $\mathbf{B}_U$.
As with the DL, each receive sub-array also experiences MUI from other UL users, which is described by the channel vectors $\mathbf{h}_{U,\tilde{j}}$ between the BS and UL user $\tilde{j}$, where $\tilde{j}\ne j$. 
The last term accounts for combined noise after analog combining and baseband processing, with $\mathbf{w}_U\sim\mathcal{CN}(\mathbf{0},\sigma^2\mathbf{I}_{M_U})$.

The third term captures the residual SI caused by the BS’s own DL transmissions leaking into its UL Rx, an inherent challenge in FD systems due to the close proximity of the co-located Tx and Rx antennas.
The SI channel matrix is denoted as $\mathbf{H}_{\mathrm{SI}}\in\mathbb{C}^{M_{U}\times M_{D}}$, which encompasses both near-field coupling and far-field NLoS components from environmental scatterers.
In the SC-HBF architecture, $\mathbf{H}_{\mathrm{SI}}$ can be structured as a block matrix composed of sub-matrices $\tilde{\mathbf{H}}_{\mathrm{SI},\{p_i,q_j\}} \in \mathbb{C}^{M_{us}\times M_{ds}}$, where each block represents the SI channel between the Rx sub-array $q_j$ that is used to serve the $j^\text{th}$ UL user and the DL sub-array $p_i$ which is used to serve the $i^\text{th}$ DL user at the BS.
Thus, the $\mathbf{H}_{\mathrm{SI}}$ can be expressed as
\begin{equation}
\begin{split}
&\mathbf{H}_{\text{SI}}= \\&\begin{bmatrix}
\tilde{\mathbf{H}}_{\mathrm{SI},\{p_1,q_1\}} & \tilde{\mathbf{H}}_{\mathrm{SI},\{p_2,q_1\}}         & \cdots & \tilde{\mathbf{H}}_{\mathrm{SI},\{p_{N_D},q_{1}\}}          \\
\tilde{\mathbf{H}}_{\mathrm{SI},\{p_1,q_2\}} &\tilde{\mathbf{H}}_{\mathrm{SI},\{p_2,q_2\}} & \cdots & \tilde{\mathbf{H}}_{\mathrm{SI},\{p_{N_D},q_{2}\}}          \\
\vdots     & \vdots         & \ddots & \vdots     \\
\tilde{\mathbf{H}}_{\mathrm{SI},\{p_{1},q_{N_U}\}} &\tilde{\mathbf{H}}_{\mathrm{SI},\{p_{2},q_{N_U}\}}              & \cdots &\tilde{\mathbf{H}}_{\mathrm{SI},\{p_{N_D},q_{N_U}\}}
\end{bmatrix}. \label{SI_matrix}
\end{split}
\end{equation}

Based on the spatial location channel model \cite{cheng2014communicating}, 
the channel vector between the BS array (operating in either UL or DL) and a given $i^\text{th}$ user 
is expressed as the superposition of signals arriving via multiple propagation paths
\begin{equation}\label{eq:intended_channel}
\mathbf{h}_i
=\sum_{l=1}^{L}
\tau_{il}^{-\eta}\,z_{il}\,\boldsymbol{g}_i(\theta_{il},\psi_{il})
=\boldsymbol{G}_i \, \mathbf{z}_i
\;\in\;\mathbb{C}^{M_i \times 1},
\end{equation}
where \(\tau_{il}^{-\eta}\) denotes the distance-dependent attenuation of the \(l\)-th path, 
\(z_{il}\sim\mathcal{CN}\!\left(0,\tfrac{1}{L}\right)\) represents its complex path gain, 
and \(\eta\) is the path-loss exponent. 
Here, \(M_i\in\{M_{D},M_{U}\}\) denotes the number of Tx and Rx antenna elements within the array. 
The vector \(\boldsymbol{g}_i(\theta,\psi)\in\mathbb{C}^{M_i\times1}\) 
characterizes the array response, while $\mathbf{z}_i=\bigl[\tau_{i1}^{-\eta}z_{i1},\dots,\tau_{iL}^{-\eta}z_{iL}\bigr]^T \in \mathbb{C}^{L \times 1}$ collects the effective path gains, and $\boldsymbol{G}_i=\bigl[\boldsymbol{g}_i(\theta_{i1},\psi_{i1}),\dots, \boldsymbol{g}_i(\theta_{iL},\psi_{iL})\bigr]\in\mathbb{C}^{M_i\times L}$ collects the response vectors across all \(L\) paths.

Since the beam resolution and the ability to achieve beam focusing and nulling 
are primarily determined by the number of array elements in the considered plane, 
a uniform linear array (ULA) sub-array configuration with effective steering angle \(\psi\) is adopted with a fixed $\theta$ angle. 
This choice streamlines the system model and facilitates direct comparison between the array aperture size and the achievable isolation performance.

\section{Joint SAS and HBF Optimization}\label{Joint_SAS_HBF}
To guarantee reliable FD mMIMO performance, a single stage of SI isolation or cancellation is insufficient to suppress SI to the noise floor and fully eliminate its detrimental impact.
This limitation arises from several practical factors:
1) the prohibitive hardware and computational complexity associated with implementing a single-stage SIC process;
2) non-linear distortions introduced at the front-end ADC or LNA caused by excessive SI power; and
3) the inadequate RF isolation provided by a single-stage antenna design or isolation-enhancement structure.

Although the deployment of large-scale antenna arrays in FD communications increases hardware and signal processing complexity, it simultaneously enables the design of multi-stage SI isolation and cancellation architectures. In FD MU-mMIMO systems, large-scale arrays provide an effective means to enhance desired signal power while suppressing interference. In particular, spatial isolation can be reinforced through two complementary strategies:
\begin{itemize}
    \item \textbf{Directional beamforming:} By shaping the array radiation pattern, energy is concentrated toward intended users, thereby minimizing spillover into unintended regions and reducing interference leakage.  
    \item \textbf{Interference nulling:} 
    By steering the beam pattern to place nulls in the directions of interference, the array’s sensitivity to unwanted signals is significantly reduced.
\end{itemize}

\subsection{Null Space Projection and MUI Suppression}
To effectively mitigate interference in MU operations, both DL and UL beamformers play a critical role. 
Interference arises when signals intended for one user cause undesired power leakage toward other users, thereby degrading the SINR. 
To address this challenge, the BS can strategically introduce spatial nulls in its beam patterns.
Specifically, the DL transmit beamforming vector $\mathbf{f}_D$ is designed to direct minimal power toward unintended users, while the UL receive beamforming vector $\mathbf{f}_U$ is shaped to suppress interfering signals by placing directional nulls in their arrival angles.

Since the considered SC-HBF architecture employs analog RF beamformers implemented as constant-modulus steering vectors, the RF beamforming vectors are explicitly determined by their steering directions. Specifically, for the $i^\text{th}$ DL sub-array and the $j^\text{th}$ UL sub-array, the RF beamforming vectors are defined as
\begin{equation}
\mathbf{f}_{D}(\psi_{D_i})=
\frac{1}{\sqrt{M_{ds}}}[1, e^{jkd\cos(\psi_{D_i})}, ...,e^{jkd(M_{ds}-1)\cos(\psi_{D_i})}]^{T},
\end{equation}
\begin{equation}
\mathbf{f}_{U}(\psi_{U_j})=
\frac{1}{\sqrt{M_{us}}}
[1, e^{jkd\cos(\psi_{U_j})},..., e^{jkd(M_{us}-1)\cos(\psi_{U_j})}]^{T},
\end{equation}
where $k$ is the wavenumber, and $d$ is the inter-element spacing.

For the $i^\text{th}$ DL user, the leakage matrix $\mathbf{\tilde{A}}_{D_i}\in \mathbb{C}^{(K_D-1) \times M_{ds}}$ collects the responses of the $i^\text{th}$ DL sub-array toward all other DL users located in the potential interference user directions $[\psi_{D_1},\dots,\psi_{D_{i-1}},\psi_{D_{i+1}},\dots,\psi_{D_{K_D}}]$.
Similarly, for the $j^\text{th}$ UL user, the leakage matrix $\mathbf{\tilde{A}}_{U_j}\in \mathbb{C}^{(K_U-1) \times M_{us}}$ collects the responses of the $j^\text{th}$ UL sub-array corresponding to all other UL-user directions $[\psi_{U_1},\dots,\psi_{U_{j-1}},\psi_{U_{j+1}},\dots,\psi_{U_{K_U}}]$. 
The columns of $\mathbf{\tilde{A}}_{D_i}^{H}$ and $\mathbf{\tilde{A}}_{U_j}^{H}$ characterize the interfering steering responses, and their column spaces therefore define the corresponding interference subspaces. To identify the associated null spaces, singular value decomposition (SVD) \cite{van2002optimum} can be applied as
\begin{equation}
\mathbf{\tilde{A}}_{D_i}
=
\mathbf{\tilde{U}}_{D_i}\mathbf{\tilde{\Sigma}}_{D_i}
\begin{bmatrix}
\mathbf{\tilde{V}}_{D_i}^{(1)} & \mathbf{\tilde{V}}_{D_i}^{(0)}
\end{bmatrix}^{H},
\end{equation}
\begin{equation}
\mathbf{\tilde{A}}_{U_j}
=
\mathbf{\tilde{U}}_{U_j}\mathbf{\tilde{\Sigma}}_{U_j}
\begin{bmatrix}
\mathbf{\tilde{V}}_{U_j}^{(1)} & \mathbf{\tilde{V}}_{U_j}^{(0)}
\end{bmatrix}^{H},
\end{equation}
where $\gamma_{D_i}=\mathrm{rank}(\mathbf{\tilde{A}}_{D_i})\leq \min(K_D-1,M_{ds})$ and $\gamma_{U_j}=\mathrm{rank}(\mathbf{\tilde{A}}_{U_j})\leq \min(K_U-1,M_{us})$. Here, $\mathbf{\tilde{V}}_{D_i}^{(1)}\in\mathbb{C}^{M_{ds}\times\gamma_{D_i}}$ and $\mathbf{\tilde{V}}_{U_j}^{(1)}\in\mathbb{C}^{M_{us}\times\gamma_{U_j}}$ span the corresponding interference subspaces, while $\mathbf{\tilde{V}}_{D_i}^{(0)}\in\mathbb{C}^{M_{ds}\times(M_{ds}-\gamma_{D_i})}$ and $\mathbf{\tilde{V}}_{U_j}^{(0)}\in\mathbb{C}^{M_{us}\times(M_{us}-\gamma_{U_j})}$ span their null spaces.

If unconstrained RF beamformers were allowed and the corresponding null spaces were nonempty, exact nulling could be achieved by selecting beamforming vectors directly from these null spaces. 
However, with the constant-modulus RF beamformers, instead of designing from the exact null space, the proposed design determines $\hat{\psi}_{D_i}$ and $\hat{\psi}_{U_j}$ such that the corresponding steering vectors maximize their projections onto the null spaces. Therefore, the orthogonal projection matrices can be defined as
\begin{equation}
\mathbf{P}_{D_i}
=
\mathbf{\tilde{V}}_{D_i}^{(0)}
\bigl(\mathbf{\tilde{V}}_{D_i}^{(0)}\bigr)^H
\in\mathbb{C}^{M_{ds}\times M_{ds}},
\end{equation}
\begin{equation}
\mathbf{P}_{U_j}
=
\mathbf{\tilde{V}}_{U_j}^{(0)}
\bigl(\mathbf{\tilde{V}}_{U_j}^{(0)}\bigr)^H
\in\mathbb{C}^{M_{us}\times M_{us}}.
\end{equation}
Accordingly, the MUI-suppression beamformer design can be formulated as
\begin{equation}
\max_{\hat{\psi}_{D_i},\,\hat{\psi}_{U_j}} \quad
\mathbf{f}_{D}^{H}(\hat{\psi}_{D_i})\mathbf{P}_{D_i}\mathbf{f}_{D}(\hat{\psi}_{D_i})+
\mathbf{f}_{U}^{H}(\hat{\psi}_{U_j})\mathbf{P}_{U_j}\mathbf{f}_{U}(\hat{\psi}_{U_j}),
\label{eq:optimization_problem_MUI}
\end{equation}
where the objective quantifies the alignment of the DL and UL RF steering vectors with the corresponding interference-nulling subspaces. In this way, the proposed design preserves the constant-modulus structure of the analog beamformers while maximizing their interference-suppression capability.

\subsection{SAS for SI Mitigation}
The abundance of antenna elements in large-scale arrays provides additional degrees of freedom for SI suppression. 
In particular, an optimal DL and UL sub-array pair $\{p_i,q_j\}$ can be selected based on user locations and SI-channel characteristics. 
Antenna configurations that reduce coupling between Tx and Rx elements minimize the effective SI channel. 
The residual SI power between the selected sub-arrays after processing with the UL and DL beamformers is given by
\begin{equation} \label{Subarray_SI}
\bigl|\mathbf{f}_U^H(\hat\psi_{U_j})\,\tilde{\mathbf{H}}_{\mathrm{SI},\{p_i,q_j\}}\,\mathbf{f}_D(\hat\psi_{D_i})\bigr|^2.
\end{equation} 

\subsection{Joint MUI and SI Suppression} 
To enable efficient MU-FD operation in the proposed HBF-mMIMO system, both the beam perturbation angles $(\hat\psi_{D_i},\hat\psi_{U_j})$ and the sub-array assignments $(p_i,q_j)$ are jointly optimized to suppress SI and MUI, as formulated in (\ref{eq:optimization_problem_MUI}) and (\ref{Subarray_SI}).
By slightly perturbing the beam directions and adaptively selecting sub-arrays, a minor loss in desired directivity can be exchanged for improved alignment of nulls toward strong MUI and SI, thereby achieving substantial gains in interference suppression.

For the scenario with $K_U$ UL and $K_D$ DL users, the joint optimization vector $\mathbf{\hat{X}}$ is expressed as
\begin{equation}
\begin{aligned}
\mathbf{\hat{X}} = [\hat{\psi}_{D_1},\hat{\psi}_{D_2},&\dots,\hat{\psi}_{D_{K_D}},\hat{\psi}_{U_1},\hat{\psi}_{U_2},\dots,\hat{\psi}_{U_{K_U}}, \\ &\hat{p}_1, \hat{p}_2,\dots,\hat{p}_{K_D}, \hat{q}_1, \hat{q}_2,\dots,\hat{q}_{K_U}]^T,
\end{aligned}\label{eq:PSO_2}
\end{equation}
where $\hat{\psi}_{D_i}$ and $\hat{\psi}_{U_j}$ denote the perturbed beam steering angles for the $i^\text{th}$ DL and $j^\text{th}$ UL user, while $\hat{p}_i$ and $\hat{q}_j$ represent the selected DL and UL sub-array indices, respectively.  

The objective function for joint SI and MUI power suppression under a combination of angular perturbation-based null-space projection beamforming  and sub-array selection is formulated as
\begin{equation}
\begin{aligned}
\min_{\mathbf{\hat{X}}} \mathcal{L}(\mathbf{\hat{X}})&= \min_{\mathbf{\hat{X}}}  \sum_{i=1}^{K_D} \sum_{j=1}^{K_U} \left|  \mathbf{f}_{U}^H(\hat{\psi}_{U_j}) \tilde{\mathbf{H}}_{\mathrm{SI},\{\hat{p}_i,\hat{q}_j\}} \mathbf{f}_D(\hat{\psi}_{D_i}) \right|^2 \\
-&\zeta_1\sum_{i=1}^{K_D}\Bigg(\mathbf{f}_{D}^H(\hat{\psi}_{D_i}) \mathbf{\tilde{V}}_{D_i}^{(0)}  
\left( \mathbf{\tilde{V}}_{D_i}^{(0)} \right)^H \mathbf{f}_{D}(\hat{\psi}_{D_i})\Bigg)^2
\\- &\zeta_2\sum_{j=1}^{K_U}\Bigg(\mathbf{f}_{U}^H(\hat{\psi}_{U_j}) \mathbf{\tilde{V}}_{U_j}^{(0)} 
\left( \mathbf{\tilde{V}}_{U_j}^{(0)} \right)^H \mathbf{f}_{U}(\hat{\psi}_{U_j}) \Bigg)^2 
\\
\text{s.t.}\; & \;\mathcal{C}_1 \;:\; \|\mathbf{f}_{D}(\hat{\psi}_{D_i})\|_2 = 1, i=1,\ldots,K_D,\qquad\\
 &\; \mathcal{C}_2 \;:\;\|\mathbf{f}_{U}(\hat{\psi}_{U_j})\|_2 = 1,\; j=1,\ldots,K_U, \\
& \;\mathcal{C}_3 \;:\;1 -\bigl|\mathbf{f}_D^H(\hat\psi_{D_i})\,\mathbf{a}_{D,i}(\psi_{D_i})\bigr|^2 \le \epsilon_i, \\
&\; \mathcal{C}_4 \;:\;1 -\bigl|\mathbf{f}_U^H(\hat\psi_{U_j})\,\mathbf{a}_{U,j}(\psi_{U_j})\bigr|^2 \le \epsilon_j,\\
&\; \mathcal{C}_5 \;:\; \hat p_i \neq \hat p_{i'} , \forall\,i\neq i';
\quad \hat q_j\neq \hat q_{j'},\forall\,j\neq j'.
\end{aligned}
\label{eq:objective_function}
\end{equation}
Here, $\zeta_1$ and $\zeta_2$ represent the positive scaling factors which balance the MUI and SI in the optimization.
$\mathcal{C}_1$ and $\mathcal{C}_2$ enforce unit-norm constraints on the DL and UL beamformers to maintain normalized power. 
Constraints $\mathcal{C}_3$ and $\mathcal{C}_4$ limit the deviation between the original beam directions $(\psi_{D_i}, \psi_{U_j})$ and their perturbed counterparts $(\hat{\psi}_{D_i}, \hat{\psi}_{U_j})$, with tolerances $\epsilon_i$ and $\epsilon_j$, thereby preserving sufficient directivity while enabling interference suppression.
$\mathbf{a}_{D,i}(\psi)\in\mathbb{C}^{M_{ds}\times 1}$ and $\mathbf{a}_{U,j}(\psi)\in\mathbb{C}^{M_{us}\times 1}$ denote the DL/UL normalized sub-array steering vectors.
$\epsilon_i$ and $\epsilon_j$ are determined by the maximum allowable degradation in desired signal gain. In our setup, for instance, a maximum degradation of 3 dB is permitted, which corresponds to a beam angle perturbation confined within the 3 dB beamwidth range.
Finally, $\mathcal{C}_5$ ensures exclusive sub-array assignments, preventing reuse conflicts and fully exploiting the spatial degrees of freedom offered by the large-scale array.

In problem (\ref{eq:objective_function}), with $K_D$ DL users and $K_U$ UL users, a total of $2K_D + 2K_U$ variables must be optimized for ULA configurations (or $3K_D + 3K_U$ for a URA). 
An exhaustive search over all possible DL/UL sub-array choices (\( \Gamma_1 \)) and perturbed angle options (\( \Gamma_2 \)) would require \( \Gamma_1^{(K_D+K_U)} \Gamma_2^{(K_D+K_U)} \) iterations, which is computationally prohibitive.  

Moreover, the optimization problem is mixed discrete-continuous: the sub-array indices are discrete decision variables subject to the exclusivity constraint, whereas the perturbation angles are continuous. In alternating optimization (AO) or other coordinate-descent-type methods, handling this mixed-variable structure and enforcing exclusive sub-array assignment typically requires either a combinatorial assignment step or a relaxation-and-rounding procedure, both of which may introduce additional suboptimality and complicate convergence. In addition, the objective function is highly non-convex and potentially non-smooth due to the coupled effects of SI and MUI suppression, sub-array selection, null-space projection, constant-modulus RF beamforming constraints, and realistic SI channels. Consequently, gradient-based or convex-surrogate-based updates are generally less reliable.

To overcome these challenges, PSO \cite{kennedy1995particle,shi1998modified,clerc2002particle} is adopted because it performs a joint, derivative-free search over the full optimization vector and naturally accommodates the mixed discrete-continuous structure of the problem. Specifically, the discrete sub-array indices are represented by continuous particle coordinates and mapped to valid integer indices during fitness evaluation through projection, followed by a feasibility-repair step to enforce the exclusivity constraint. Furthermore, since the fitness evaluations of different particles are mutually independent, they can be carried out in parallel, making PSO attractive for practical implementation and enabling efficient use of parallel computing resources.

PSO operates with a swarm of \( N_z \) particles, each representing a candidate variable vector, which collaboratively explores the search space over \( N_t \) iterations within predefined boundaries to optimize the objective function. 
The resulting computational complexity is reduced to $\mathcal{O}(N_z N_t K)$.  
The particle updates are driven by two factors: their individual best-known positions and the swarm’s global best-known position. 
This stochastic search mechanism enhances robustness, making PSO well-suited for solving non-convex optimization problems. 
Moreover, PSO’s scalability allows the number of particles and iterations to be flexibly adjusted, offering a practical trade-off between optimization accuracy and computational cost.  
An iteration threshold can also be applied to terminate the process if no improvement is observed in the global best solution over a specified number of consecutive iterations.  

The candidate vector of the $z^\text{th}$ particle at the $t^\text{th}$ iteration is expressed as
\begin{equation}
\begin{aligned}
\mathbf{X}_z^{(t)}\hspace{-0.5ex} =& [{\psi}_{D_1}(t,z),\dots,{\psi}_{D_{K_D}}(t,z),{\psi}_{U_1}(t,z),\dots,{\psi}_{U_{K_U}}(t,z), \\ 
& {p}_1(t,z),\dots,{p}_{K_D}(t,z), {q}_1(t,z),\dots,{q}_{K_U}(t,z)]^T,
\end{aligned}\label{eq:PSO2}
\end{equation}
where $z = 1, \dots, N_z$ and $t = 0, 1, \dots, N_t$.  
For each particle, the objective function (\ref{eq:objective_function}) is evaluated as $\mathcal{L}(\mathbf{X}_z^{(t)})$.  

At the $t^\text{th}$ iteration, the personal best of the $z^\text{th}$ particle and the global best across all particles are identified as
\begin{equation}
\mathbf{X}^{(t)}_{\text{best}, z} = \arg \min_{\mathbf{X}^{(t^*)}_z} \mathcal{L}(\mathbf{X}^{(t^*)}_z), \quad \forall t^* = 0, 1, \ldots, t,
\label{eq:personal_best_PSOLPA2}
\end{equation}
\begin{equation}
\mathbf{X}^{(t)}_{\text{best}} = \arg \min_{\mathbf{X}^{(t)}_{\text{best}, z}} \mathcal{L}(\mathbf{X}^{(t)}_{\text{best}, z}), \quad \forall z = 1, 2, \ldots, N_z.
\label{eq:global_best_PSOLPA2}
\end{equation}

The effectiveness of SAS-based SI and MUI suppression depends on the velocity vector \( \mathbf{v}_z \), which incorporates both the particle’s personal best $\mathbf{X}_{\text{best}, z}$ and the global best $\mathbf{X}_{\text{best}}$.  
The velocity of the $z^\text{th}$ particle at iteration $t+1$ is updated as
\begin{equation}
\mathbf{v}^{(t+1)}_z = \mathbf{\Omega}_1 \bigl(\mathbf{X}^{(t)}_{\text{best}} - \mathbf{X}^{(t)}_z\bigr) 
+ \mathbf{\Omega}_2 \bigl(\mathbf{X}^{(t)}_{\text{best}, z} - \mathbf{X}^{(t)}_z\bigr) 
+ \mathbf{\Omega}_3^{(t)} \mathbf{v}^{(t)}_z,
\label{eq:velocity2}
\end{equation}
where \( \mathbf{v}^{(t)}_z \) denotes the velocity at iteration $t$.  
\( \mathbf{\Omega}_1 \) and \( \mathbf{\Omega}_2 \) are random diagonal matrices with entries uniformly distributed in \([0,2]\), modelling social interactions among particles and the attraction toward personal bests, respectively.  
The inertia weight matrix \(\mathbf{\Omega}_3^{(t)}\) balances exploration and exploitation. 
An adaptive inertia mechanism, such as a linearly decreasing schedule or a stall-counter-based adaptive scalar update, can be employed to promote larger particle movements for global exploration in the early stage of the optimization, while gradually reducing the step size to enable more stable local refinement near convergence.

\begin{algorithm}[t!]
    \caption{Proposed Joint Sub-array Selection and Beam Perturbation Optimization Algorithm} 
    \label{algo:1}
    \begin{algorithmic}[1] 
     \Statex \textbf{Input}: $\mathbf{H}_{\mathrm{SI}}$, ${\psi_U}_1$, ...,${\psi_U}_{K_U}$, ${\psi_D}_1$,..., ${\psi_D}_{K_D}$. 
        \Statex \textbf{Output}: $[\hat{\psi}_{D_1},...,\hat{\psi}_{D_{K_D}},\hat{\psi}_{U_1},...,\hat{\psi}_{U_{K_U}},$ 
        \Statex \hspace{16ex} $\hat{p}_1,...,\hat{p}_{K_D}, \hat{q}_1,...,\hat{q}_{K_U}]$
        \For{$t = 0:N_t$}
            \For{$z = 1:N_z$}
                \If{$t = 0$}
                    \State Initialize the velocity as $\mathbf{v}_{z}^{(0)} = \mathbf{0}$.
                    \State Initialize $\mathbf{X}_z^{(t)}$ uniformly in $[\mathbf{X}_{\text{Low}}, \mathbf{X}_{\text{Upp}}]$.
                \Else
                    \State Update the velocity $\mathbf{v}_{z}^{(t)}$ via (\ref{eq:velocity2}).
                    \State Update the vector $\mathbf{X}_{z}^{(t)}$ via (\ref{eq:position2}).
                \EndIf
                \State Find the personal best $\mathbf{X}_{\mathrm{best},z}^{(t)}$ via (\ref{eq:personal_best_PSOLPA2}).
            \EndFor
            \State Find the global best $\mathbf{X}_{\text{best}}^{(t)}$ as in (\ref{eq:global_best_PSOLPA2}).
        \EndFor
    \end{algorithmic}
\end{algorithm}

The particle position is then updated as
\begin{equation}
\mathbf{X}^{(t+1)}_z = \text{clip}\!\left( \mathbf{X}^{(t)}_z + \mathbf{v}^{(t+1)}_z, \, \mathbf{X}_{\text{Low}}, \, \mathbf{X}_{\text{Upp}} \right),
\label{eq:position2}
\end{equation}
where \( \mathbf{X}_{\text{Low}}, \mathbf{X}_{\text{Upp}} \in \mathbb{R}^{(2K_D + 2K_U)} \) denote the lower and upper bounds of the perturbation coefficients. 
The boundaries of the sub-array indices are determined by the minimum and maximum available indices, while the allowable perturbation angle ranges are pre-estimated from the thresholds $\epsilon_i$ and $\epsilon_j$ specified in constraints $\mathcal{C}_3$ and $\mathcal{C}_4$. 
The clipping function $\text{clip}(x, a, b) = \min(\max(x,a), b)$ is applied elementwise to enforce feasibility by constraining each decision variable within its bounds. 
Since sub-array indices are discrete, after clipping, they are projected onto the integer lattice via $\hat p_i \leftarrow \mathrm{round}(\hat p_i)$ and $\hat q_j \leftarrow \mathrm{round}(\hat q_j)$, followed by a repair step that removes duplicates and enforces $\mathcal{C}_5$ by greedily reassigning conflicting indices to the nearest available choices within bounds.

Finally, the RF beamforming matrices $\mathbf{F}_D$ and $\mathbf{F}_U$ can be constructed using the optimized perturbation angles for DL and UL users, i.e., $\mathbf{f}_D(\hat{\psi}_{D_i})$ for the $i^\text{th}$ DL user and $\mathbf{f}_U(\hat{\psi}_{U_j})$ for the $j^\text{th}$ UL user. The corresponding $\hat{p}_i^\text{th}$ DL and $\hat{q}_j^\text{th}$ UL sub-array are employed. The detailed steps of the algorithm are outlined in Algorithm~\ref{algo:1}.

The proposed joint SAS and APN scheme is designed as a slow-time-scale reconfiguration mechanism driven by relatively slowly varying information. In particular, the users' angular location parameters typically remain quasi-static over multiple coherence intervals under practical mobility. 
Accordingly, in the intended implementation, PSO is triggered intermittently, for example, when a user’s direction drifts beyond a predefined threshold, or when the SI coupling environment changes due to the hardware reconfiguration. 
Between two SAS/APN updates, the selected sub-arrays and RF beamformers are kept fixed, while the system can maintain real-time operation through conventional fast-time-scale baseband adaptation and updating.

\subsection{Baseband Precoder Design}
Following the design of the RF beamformers in the previous section, the baseband precoders can be determined.  
The baseband stage is constructed using the regularized zero-forcing (RZF) technique to suppress residual MUI.

Given the DL and UL RF beamformers $\mathbf{F}_{D}(\mathbf{\hat{X}})$ and $\mathbf{F}_{U}(\mathbf{\hat{X}})$, the effective reduced-dimensional channels are defined as  
\begin{equation}
    \begin{aligned}
 \mathcal{H}_D = \mathbf{H}^H_D \mathbf{F}_D \in \mathbb{C}^{K_D \times N_D}, 
\\
\mathcal{H}_U = \mathbf{F}^H_U \mathbf{H}_U \in \mathbb{C}^{N_U \times K_U},
    \end{aligned}
\end{equation}
where $\mathbf{H}_D=[\mathbf{h}_{D,1},\dots,\mathbf{h}_{D,K_D}] \in \mathbb C^{M_D\times K_D}$ and $\mathbf{H}_U=[\mathbf{h}_{U,1},\dots,\mathbf{h}_{U,K_U}]\in \mathbb C^{M_U\times K_U}$ are the channel matrices which consist of the channel from DL and UL array to all the users.

The DL baseband precoder is obtained as
\begin{equation}
\begin{aligned}
    \mathbf{B}_D = \kappa_D \mathbf{Z}_D^{-1} \mathcal{H}_D^H \in \mathbb{C}^{N_D \times K_D},
\label{eq:bd}
\end{aligned}
\end{equation}
where $\mathbf{Z}_D = \mathcal{H}_D^H \mathcal{H}_D + \left( \frac{\sigma_w^2}{P_D/K_D} \right)\mathbf{I}_{N_D} \in \mathbb{C}^{N_D \times N_D}$ and $P_D$ is total DL Tx power.

Here, $\mathbf{Z}_D$ incorporates noise power $\sigma^2$ for regularization to mitigate MUI.  
The normalization factor is given by
\begin{equation}
\begin{aligned}
\kappa_D=\sqrt{\frac{P_D}{\mathrm{tr}\!\bigl(\mathbf{F}_D\mathbf Z_D^{-1}\mathcal H_D^{H}\mathcal H_D\mathbf Z_D^{-1} \mathbf{F}^H_D \bigr)}},
\end{aligned}
\end{equation}
which ensures compliance with the DL power constraint, i.e., $\mathbb{E}\{\|\mathbf{s}_D\|^2\} = \mathrm{tr}(\mathbf{F}_D \mathbf{B}_D \mathbf{B}_D^H \mathbf{F}_D^H) = P_D$, as defined in the system model.  

Similarly, the UL baseband combiner is expressed as
\begin{equation}
\begin{aligned}
\mathbf{B}_U = \mathbf{Z}_U^{-1} \mathcal{H}_U \in \mathbb{C}^{N_U \times K_U}, \label{eq:bu}
\end{aligned}
\end{equation}
where $\mathbf{Z}_U = \mathcal{H}_U \mathcal{H}_U^H + \left( \frac{\sigma_w^2}{P_U} \right)\mathbf{I}_{N_U}
\in \mathbb{C}^{N_U \times N_U}$, with \(\alpha_U=\frac{\sigma_w^2}{P_U}\) where \(P_U\) is the per UL user power.

\section{Experimentally Measured SI Channel of FD Large-Scale Antenna Array}\label{Meas_SI}
As described in the considered FD-mMIMO system model, the following two classes of channels are involved and are treated differently to ensure both modelling accuracy and practical realism:

1) UL/DL user channels: Consistent with standard mMIMO simulation practice, the BS-user channels are modelled using a conventional far-field geometric multipath scheme formed by the superposition of multiple propagation paths.   

2) SI channel: In contrast, the SI channel is dominated by near-field MC, surface-wave leakage, and spatial leakage between the co-located Tx and Rx arrays. Owing to the close proximity of the Tx and Rx arrays and the multiple leakage mechanisms involved, the SI channel is considerably more difficult to model accurately. In addition, practical RF-isolation techniques employed in FD arrays further shape the SI response. As a result, purely geometry-based SI models are generally insufficient to accurately represent the realistic SI environment and may not faithfully capture the SI conditions targeted by the proposed SI-suppression design. Therefore, in this work, the SI channel is directly obtained from experimental measurements.

To seamlessly incorporate the experimentally measured SI channel into the simulations, a strict relative geometric and structural mapping is enforced between the physical array prototype and the simulation setup.
During post-processing of the measured SI channel data, the insertion losses introduced by the measurement setup are mathematically de-embedded so that the pure Tx-Rx antenna level MC can be extracted.

\subsection{$8\times8$Tx-$8\times8$Rx Cross-Polarized Array Prototype}
\begin{figure}[!t]
\centerline{\includegraphics[width=\linewidth]{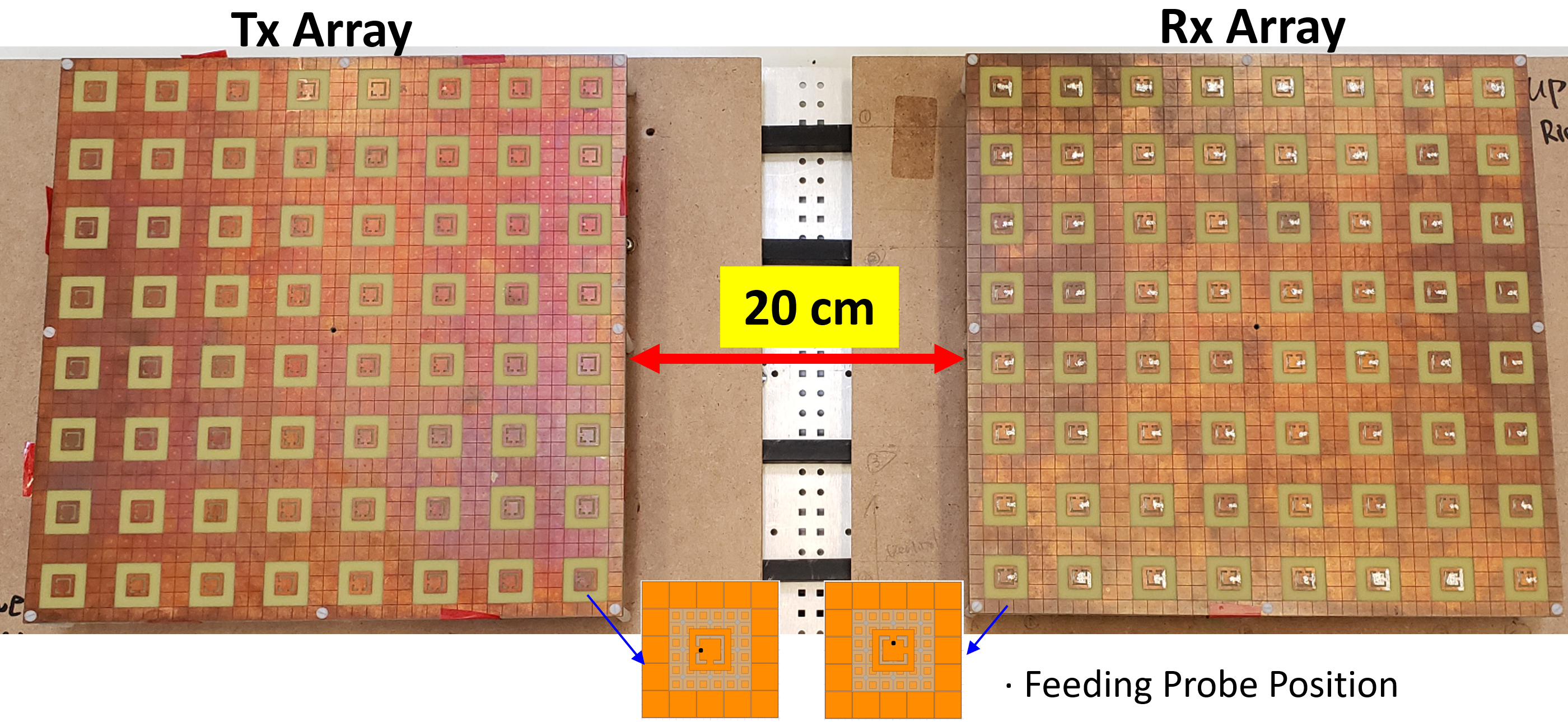}}
\caption{$8\times8$Tx-$8\times8$Rx array prototype for FD mMIMO.}
\label{FD_prototype_ch6}
\end{figure}

As shown in Fig.~\ref{FD_prototype_ch6}, an $8\times8$Tx-$8\times8$Rx antenna array prototype incorporating cross-polarized Tx-Rx antenna elements is employed in the experiments to obtain measurement-based SI channels for FD communications. 
The antenna array is designed to operate within a target bandwidth of 3.35 GHz to 3.6 GHz \cite{YG_ACCESS_1}.
Within Tx and Rx arrays, an inter-element spacing of 4.0 cm ($\approx0.5\lambda_0$ at 3.5 GHz) center-to-center is employed to avoid spatial aliasing and the formation of grating lobes.
All elements within each array share a common FR-4 dielectric substrate and ground plane. Each array measures 32 cm $\times$ 32 cm, with a 20 cm separation between the Tx and Rx arrays. Including this separation, the overall Tx-Rx prototype occupies 84 cm $\times$ 32 cm and accommodates 64 transmitting and 64 receiving elements. 
On the back side of the prototype, each antenna element is soldered with a JSC coaxial connector, which is subsequently connected to either the beamforming network (BFN) components or RF switches, depending on the measurement configuration.

\begin{figure}[!t]
\centerline{\includegraphics[width=\linewidth]{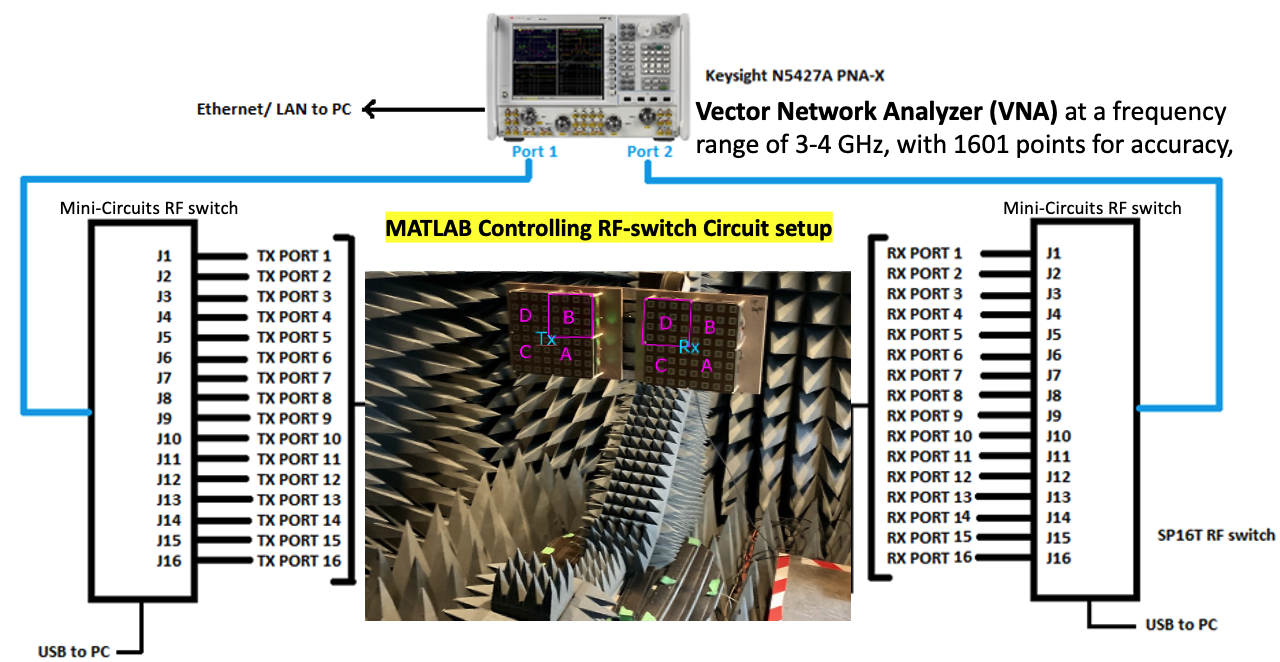}}
\caption{$8\times8$Tx-$8\times8$Rx array prototype for FD mMIMO measurement setup.}
\label{Elem_SI_meas_setup}
\end{figure}

\begin{figure}[!t]
\centerline{\includegraphics[width=\linewidth]{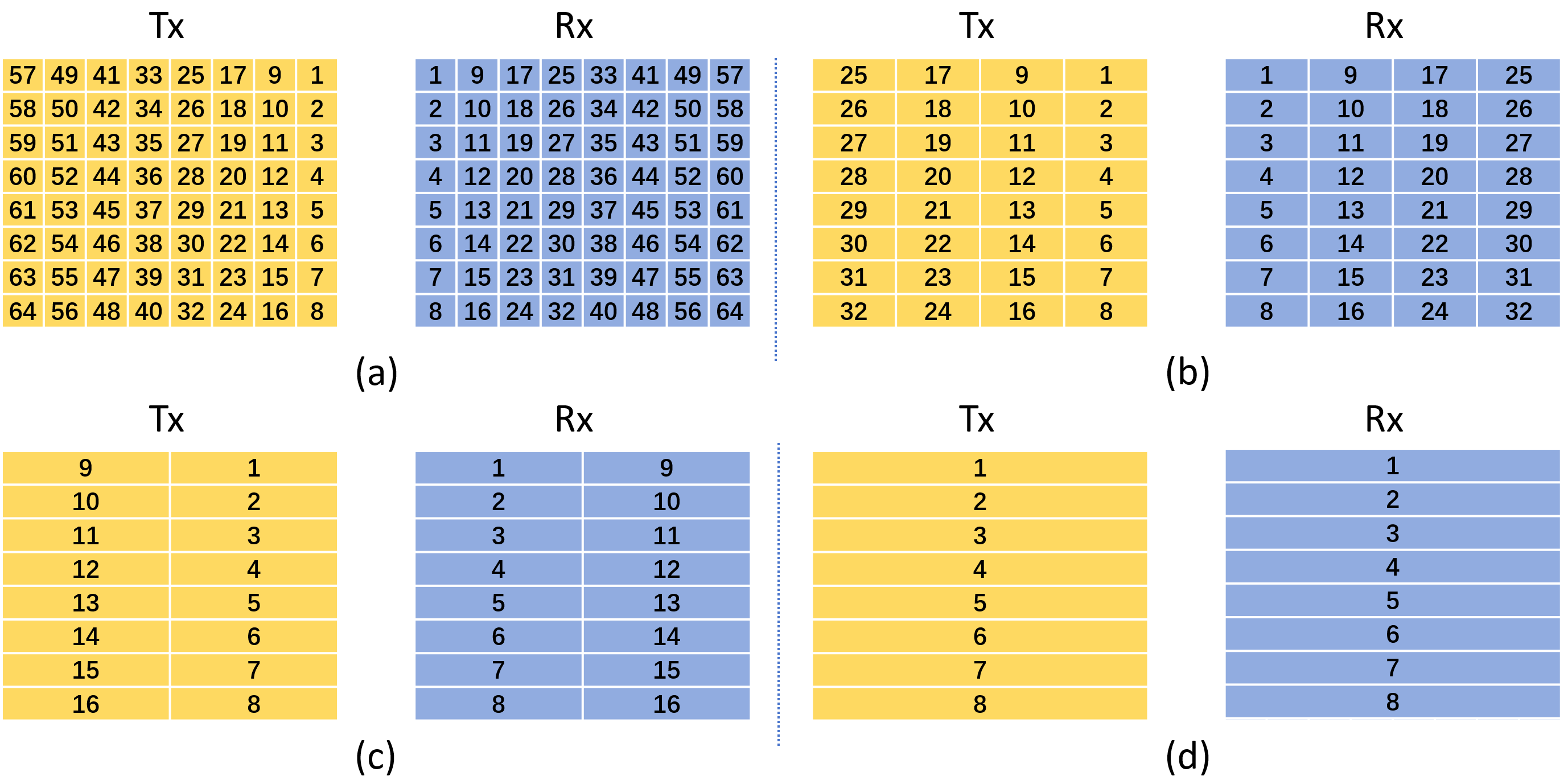}}
\caption{$8\times8$Tx-$8\times8$Rx array (a) antenna element indices; (b) 1$\times$2, (c) 1$\times$4, (d) 1$\times$8 sub-array indices.}
\label{Simul_setup}
\end{figure}

To comprehensively evaluate the MC between Tx-Rx elements and obtain realistic SI channel data, over-the-air (OTA) FD measurements were conducted in our anechoic chamber, which has internal dimensions of 20 ft (6.1 m) $\times$ 8 ft (2.4 m) $\times$ 8 ft (2.4 m) and effectively eliminates external reflections.
The array prototype was positioned more than 4 ft away from surrounding walls, each lined with C-RAM SFC-48 absorbers providing reflection levels below \(-45\) dB at 3.5 GHz.
S-parameters were measured using a Keysight N5247A PNA-X vector network analyzer (VNA) with an output power of 10 dBm. To reduce potential noise impacts in the high-isolation measurements, the intermediate frequency (IF) bandwidth was set to 300 Hz, corresponding to an integrated thermal noise power of approximately \(-149\) dBm. 
The trace post-detection smoothing function was set to 1\%, and averaging was disabled.
Measurements were performed over the 3 to 4 GHz frequency band with a step size of 625 kHz, yielding 1601 sample points. 

To automate the iterative measurement process and protect the fragile JSC connectors, two 1-to-16 Mini-Circuits USB-1SP16T-83H RF absorptive switches were employed, controlled by MATLAB-based programs to switch between different Tx-Rx antenna-element pairs, as illustrated in Fig.~\ref{Elem_SI_meas_setup}. 
Measurements were conducted in blocks of 16 Tx by 16 Rx, enabling acquisition of the full 64$\times$64 S-parameter matrix. 
The measured data were transferred to a PC for MATLAB-based post-processing. 
For a fixed RF attenuation of 100 dB at the VNA ports, variations across frequency were observed to be within $\pm 0.75$ dB in magnitude, $\pm 3^\circ$ in phase, and $\pm 100$ ps in group delay.

The resulting S-parameters were stored as .s2p-files and subsequently imported into MATLAB, where the insertion losses from the measurement setup were de-embedded to account for connector and cable losses. The de-embedded data were then combined into .s128p-files to construct the $64 \times 64$ SI channel matrix, representing the MC between all Tx and Rx elements. The complete complex SI channel matrix, $\mathbf{H}_{\text{SI}}$, had dimensions of $64 \times 64 \times 1601$, corresponding to the 1601 measured frequency points.

\subsection{$8\times8$Tx-$8\times8$Rx Experimental SI Channel}
\begin{figure*}[!t]
\centerline{\includegraphics[width=\textwidth]{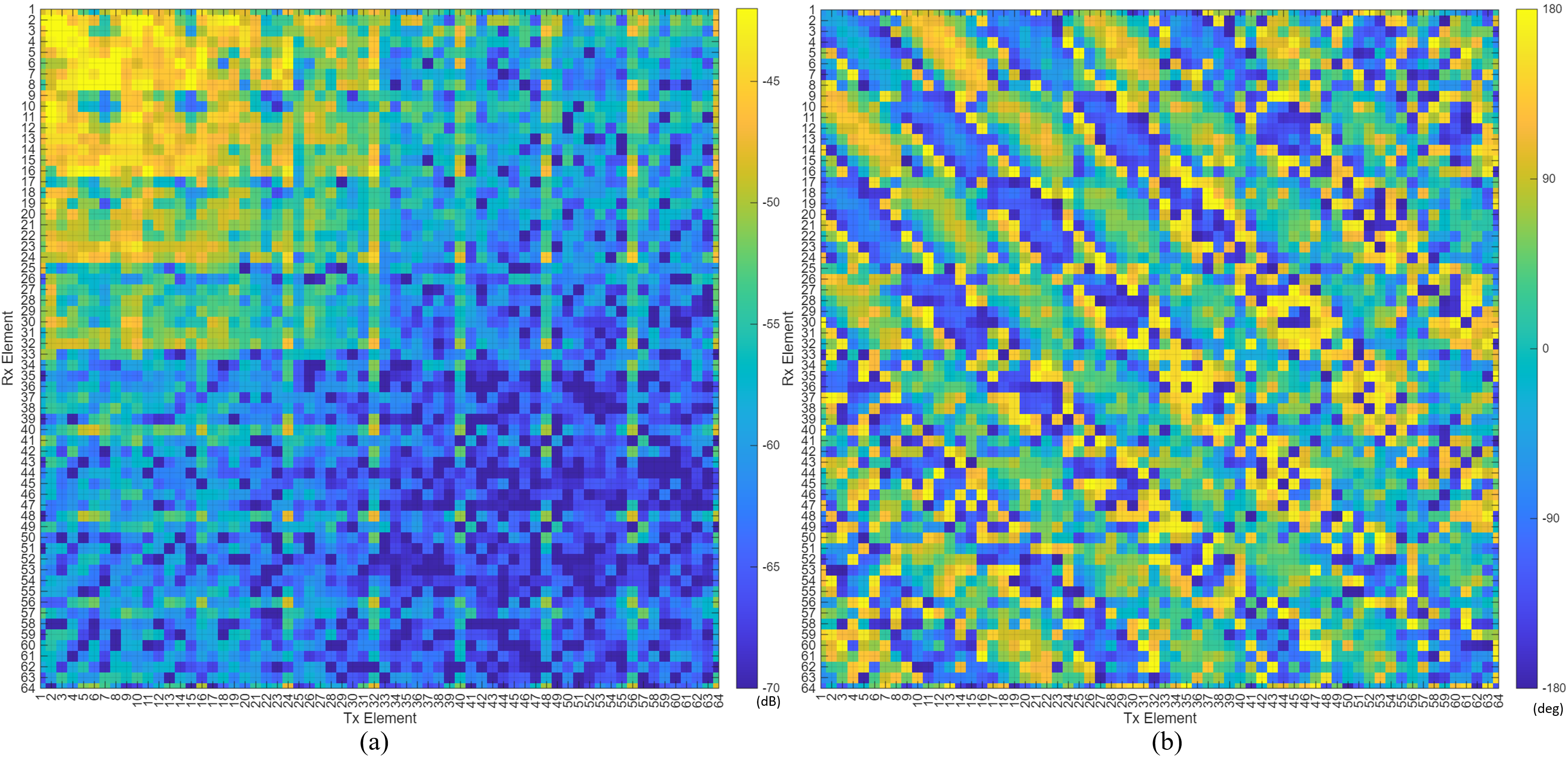}}
\caption{64 Tx element to 64 Rx element mutual coupling at 3.5 GHz (a) magnitude in dB maps; (b) phase in degree maps (measurement results).}
\label{Tx_RX_SI}
\end{figure*}

The comprehensive measured SI channel between the 64 Tx and 64 Rx elements at 3.5 GHz is illustrated in Fig.~\ref{Tx_RX_SI}. 
As presented in the magnitude plot  Fig.~\ref{Tx_RX_SI} (a), a clear dependence of MC on element spacing is observed. As the separation increases, surface-wave leakage between the elements progressively diminishes. Additionally, the inclusion of additional electromagnetic bandgap (EBG) patches between Tx and Rx elements further improves isolation by filtering and mitigating the surface leakage wave. 
The two Tx-Rx element pairs with the smallest spatial separation exhibit the strongest (worst) MC, which are: 1) Tx-8 and Rx-6, where the Tx element is located at the array corner and the Rx element at the array edge, with an MC of $-37.8$ dB; and 2) Tx-2 and Rx-2, which have the smallest separation distance, with an MC of $-38.5$ dB. 
Among the 64-by-64 antenna element pairs, the lowest (best) MC of $-101.3$ dB can be achieved. Across all 4096 Tx-Rx pairs, 99.9\% (4092 pairs) achieve an MC better than $-40$ dB, 87.0\% (3562 pairs) achieve better than $-50$ dB, 48.3\% (1980 pairs) achieve better than $-60$ dB, and 8.6\% (351 pairs) achieve better than $-70$ dB.
Tx-Rx pairs with smaller separations, such as those in the top-left quadrant of the MC heatmap, exhibit stronger SI levels.
The average MC between Tx-1 to Tx-32 and Rx-1 to Rx-32 (closer Tx-Rx element pairs) is $-51.1$ dB, whereas the average MC between Tx-33 to Tx-64 and Rx-33 to Rx-64 (bottom-right quadrant) is significantly lower at $-66.1$ dB, yielding a difference of approximately $15.0$ dB.
Among these 1024 Tx-Rx pairs, 99.4\% achieve an MC better than $-55$ dB, and 85.7\% achieve better than $-60$ dB. 
For Tx-Rx pairs with comparable relative positions, similar MC levels are observed. For example, the average MC in the top-right and bottom-left quadrants of Fig.~\ref{Tx_RX_SI}~(a) is $-59.5$ dB and $-61.4$ dB, respectively, with an average difference of only $1.9$ dB.

As presented in Fig.~\ref{Tx_RX_SI}~(b), a distinct periodic phase pattern in the Tx-Rx element pair position iteration is observed along the diagonal direction of the plot. 
This behaviour originates from variations in coupling path lengths, which shift with half-wavelength spacing between the elements. 
In particular, when the Tx-Rx separation approaches a wavelength or its integer multiples, the coupled waves exhibit a repeatable phase relationship, producing the observed periodic phase transitions across the arrays.
Out-of-phase MC components can be beneficial for enhancing spatial isolation. Their inherent destructive interference suppresses unwanted signal leakage among neighbouring elements during signal summation. 
This feature is particularly advantageous for sub-array beamforming isolation design. By exploiting out-of-phase SI cancellation, residual SI can be further mitigated, thereby improving inter-sub-array isolation, an aspect that will be further elaborated in the following sections.

\section{Illustrative Results}\label{Results}
\begin{figure}[!t]
\centerline{\includegraphics[width=\linewidth]{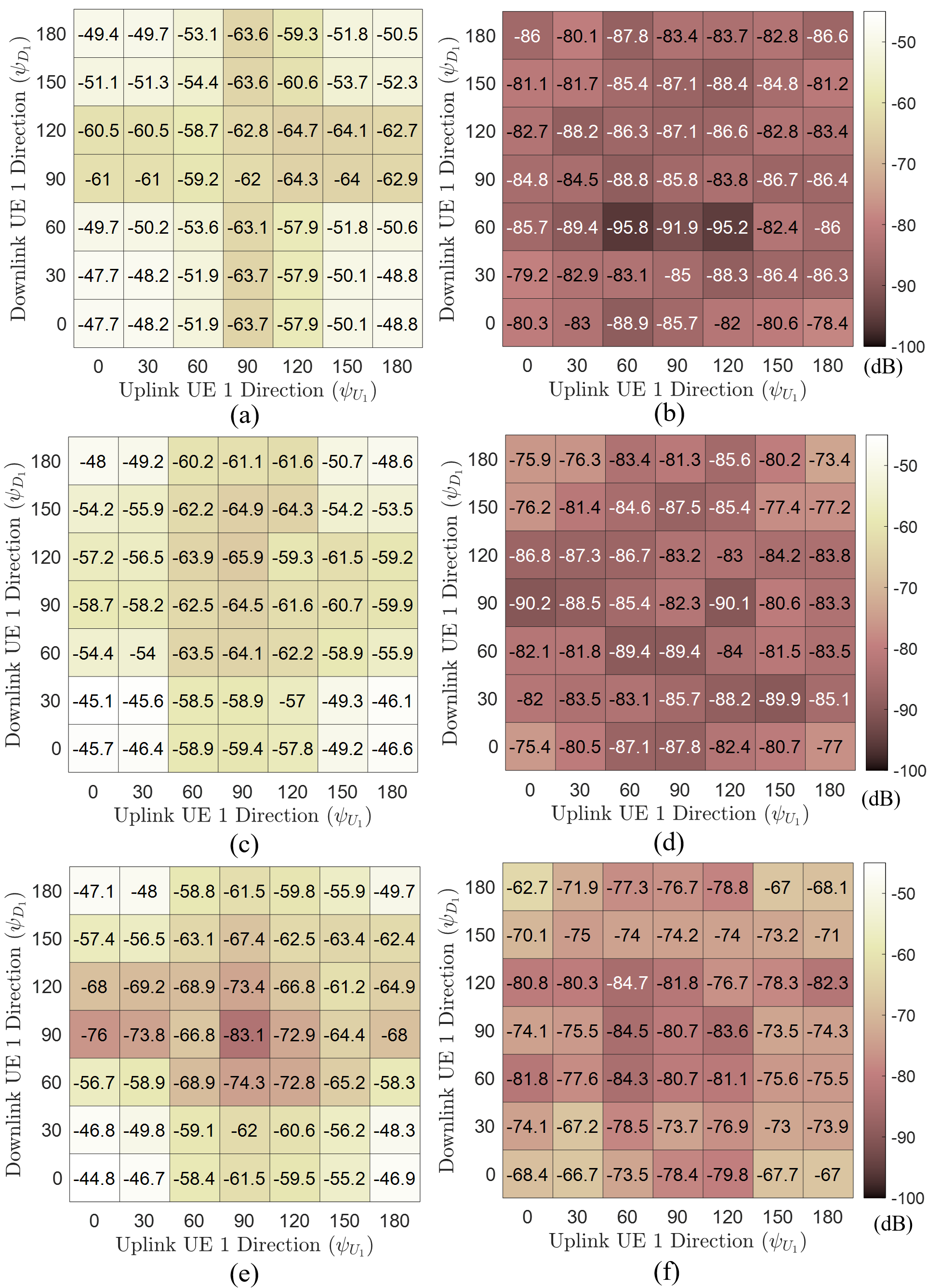}}
\vspace{-0.2cm}
\caption{Tx-Rx beam-level mutual coupling with angular perturbation-based null-space projection beamforming, for 1$\times$2 (a) without SAS, (b) with SAS; for 1$\times$4 sub-arrays (c) without SAS, (d) with SAS; for 1$\times$8 sub-arrays (e) without SAS, (f) with SAS.}
\label{SAS_SI_Imp2}
\end{figure}
\subsection{Simulation Setup}
The measurement-based SI channel $\mathbf{H}_{\mathrm{SI}}$ obtained from the $8\times8$ Tx-$8\times8$ Rx prototype, as presented in the previous section, is employed to evaluate the performance of the proposed joint SAS and HBF optimization scheme. 
To provide better insights into beam-level isolation across different user locations and to explore the flexibility offered by selective sub-array SI channels, the FD simulations consider a two-user DL and a two-user UL scenario, where $N_D = N_U = K_D = K_U = 2$.  
The key simulation parameters are summarized in Table~\ref{tab:simp1}.  
The illustrative results employ geometric means to characterize the average normalized beam-level SI and MUI, since the MC is expressed as the ratio of received to transmitted power.
\begin{table}[!t]
    \centering
    \caption{Simulation Parameters}
    \label{tab:simp1}
    \renewcommand{\arraystretch}{1}
    \small
    \begin{tabular}{|c|c|}
        \hline
        Number of antennas (DL/UL) & $M_D = M_U = 64$ \\ \hline
        Sub-array size (DL/UL) & $M_{ds} = M_{us} = 2,4,8$ \\ \hline
        Number of users (DL/UL) & $K_D = K_U = 2$ \\ \hline
        $\psi$ range / $\theta$ angle & $0^\circ:30^\circ:180^\circ$ / $90^\circ$\\ \hline
        Frequency & 3.5 GHz \\ \hline
        Number of paths & $L = 20$ \\ \hline
        Path-loss exponent & 3.76 (NLOS) \\ \hline
        User distance & 15 m \\ \hline 
        Transmit power for DL & $P_D = 10$ dBm \\ \hline
        Transmit power for UL & $P_U = 10$ dBm \\ \hline
        
    \end{tabular}
\end{table}

For the PSO setup, since the proposed SAS and APN design involves a relatively low-dimensional optimization vector, a moderate swarm size is generally sufficient to provide adequate search diversity. In practice, the swarm size is also selected in accordance with the available computational resources, such as the number of CPU cores and parallel workers, so as to balance robustness, exploration capability, and implementation efficiency. 
In this work, for a solution-vector dimension of \(n_{\mathrm{var}}\), the swarm size is set to \(50\,n_{\mathrm{var}}\).
To ensure sufficiently thorough convergence and to fully evaluate the capability of the joint optimization for SI and MUI suppression, the maximum number of iterations is set to \(200\,n_{\mathrm{var}}\). Moreover, a stall-based stopping criterion is adopted, such that the optimization terminates when the relative change in the swarm-best objective value over the most recent 20 iterations falls below \(10^{-6}\). This criterion allows the optimization to terminate automatically once further improvement becomes negligible, thereby avoiding unnecessary computation while maintaining stable convergence.
An adaptive inertia mechanism based on MATLAB’s stall-counter-based scalar update is adopted, with the inertia bounded within the predefined range \([0.1,\,1.1]\). Specifically, the stall counter is decreased when the swarm-best objective value improves and increased otherwise. Accordingly, the inertia is adaptively increased or decreased within the prescribed range, such that a small counter corresponds to a larger inertia for broader exploration, whereas a large counter yields a smaller inertia for more stable local refinement. As a result, the solver encourages global exploration in the early stage of the search while enabling more stable local refinement near convergence. 

\subsection{Sub-array Selection Scheme Improvement in SI Suppression}
In the simulations, DL/UL user~1 is assumed to vary within an angular range of $\{\psi_{D_1}, \psi_{U_1}\} \in [0^\circ,180^\circ]$, while DL/UL user~2 is fixed at the BS broadside location, i.e., $\psi_{D_2}, \psi_{U_2} = 90^\circ$.
The average beam-level MC performance obtained with the proposed DL/UL beamforming optimization schemes with and without the SAS scheme is illustrated in Fig.~\ref{SAS_SI_Imp2}, where three sub-array configurations of sizes $1\times2$, $1\times4$, and $1\times8$ are compared.  

\subsubsection{Achievable SI Level without Sub-array Selection}
When fixed DL/UL sub-arrays are employed, beam-level isolation improves as the sub-array length increases. 
This enhancement is attributed to the higher beam directivity, which provides stronger spatial isolation by suppressing interference leakage in undesired directions and reducing the angular spacing between achievable interference-nulling locations. 
For instance, increasing the sub-array length from $1\times2$ to $1\times8$ improves the average DL-UL beam-level isolation from $-56$ dB to $-61.3$ dB, yielding an average gain of 5.3 dB. Concurrently, the best-case (minimum) DL-UL MC level also improves from $-64.7$ dB to $-83.1$ dB, representing an $18.4$ dB enhancement.

Moreover, the results further reveal a strong angular dependence in isolation performance. Due to the fixed SI channel and the limited steerability of the array under beam perturbation and null-space projection algorithms, high isolation can only be achieved for specific DL/UL user angular pairs. 
Under the $1\times8$ configuration, beam-level isolation ranges from $-44.8$ dB to $-83.1$ dB, yielding a 38.4 dB variation across different angular pairs.
In the case of the $1\times2$ configuration, a performance change of 17 dB is observed between the best and the worst target user locations.
As users move away from the array broadside, beam-level isolation degrades noticeably. This degradation becomes more pronounced for smaller apertures, where reduced steerability and narrower concentration regions limit the angular range over which high DL-UL isolation can be sustained.

\subsubsection{Achievable SI Level with Sub-array Selection}
The employment of sub-array selection techniques significantly enhances the achievable DL-UL beam-level SI mitigation performance by introducing reconfigurable SI channels between sub-arrays. 
For sub-array configurations of $1\times2$, $1\times4$, and $1\times8$, the sub-array selection approach yields average improvements of $29.2$ dB, $26.6$ dB, and $14.3$ dB, respectively. 
With the proposed SI channel selection approach, average beam-level isolation levels of $-85.2$ dB, $-83.3$ dB, and $-75.5$ dB are achieved for the $1\times2$, $1\times4$, and $1\times8$ sub-array configurations, respectively.

Additionally, the worst-case DL-UL beam-level SI mitigation improvements reach $30.7$ dB, $28.3$ dB, and $17.9$ dB for the $1\times2$, $1\times4$, and $1\times8$ configurations, respectively. 
With sub-array selection, the beam-level isolation performance is consistently better than $-78.4$ dB and $-73.4$ dB for the $1\times2$ and $1\times4$ configurations, respectively.
Moreover, a more uniform high-isolation performance can be observed across all DL/UL user locations. 
These results confirm that the increased flexibility in sub-array configuration and SI channel reconfigurability substantially enhances both the robustness and overall effectiveness of the proposed approach.

Notably, the $1\times2$ sub-array configuration achieves an additional 14.9 dB improvement compared with the $1\times 8$ configuration. This gain effectively compensates for the performance degradation typically associated with reduced aperture size, such as lower beam directivity and limited steerability, by leveraging the greater flexibility in SI channel selection.

\subsubsection{Distributions of Selected Optimal Sub-array}\label{sub_indices}
\begin{figure}[!t]
\centerline{\includegraphics[width=\linewidth]{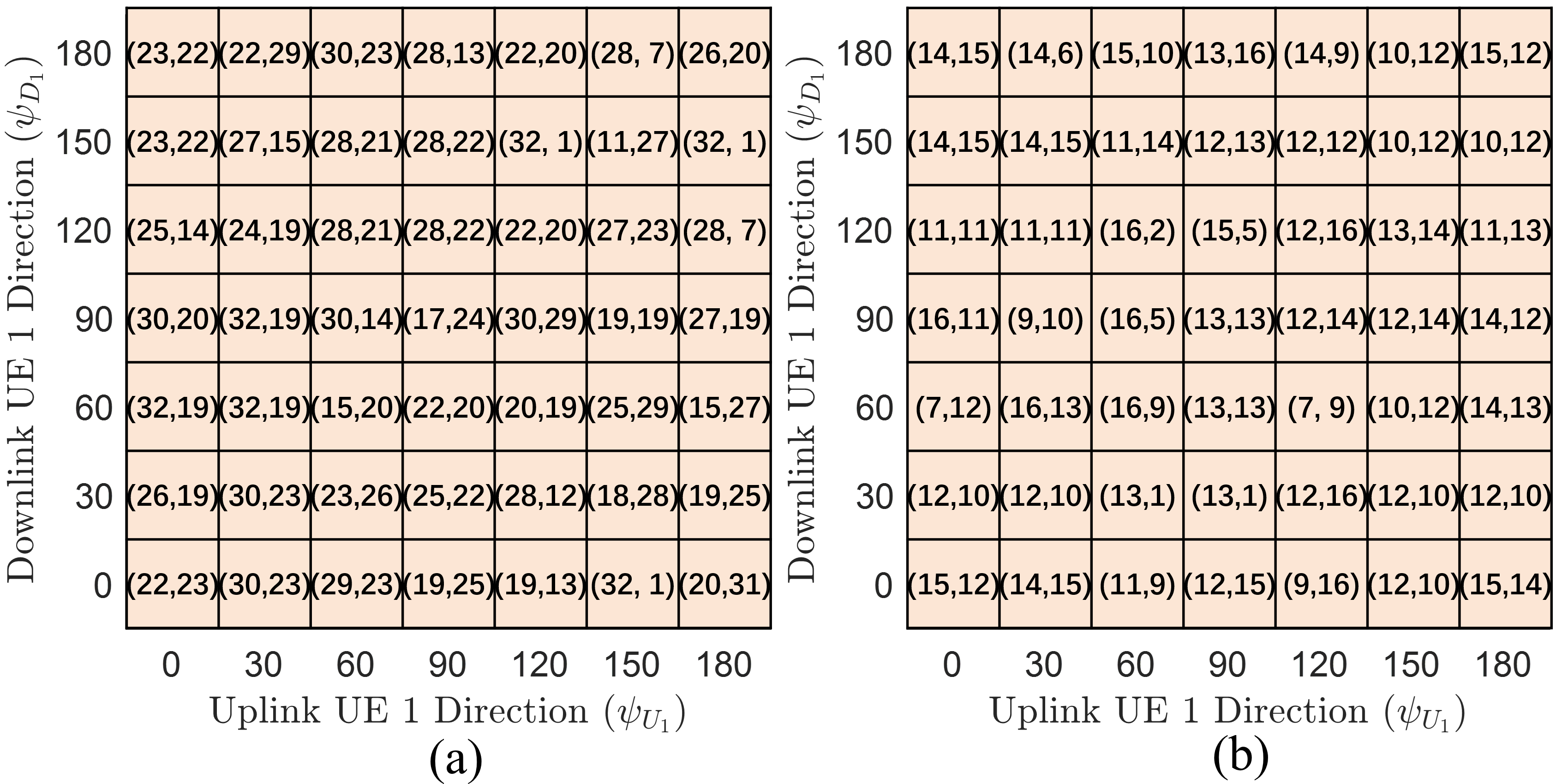}}
\vspace{-0.2cm}
\caption{DL-UL selected sub-array indices $(p_1,q_1)$ with (a) 1$\times$2 and (b) 1$\times$4 sub-array configurations.}
\label{SAS_Index}
\end{figure}
To further examine the behaviour of the sub-array selection optimization algorithm in determining the optimal DL/UL sub-arrays, the selected sub-array indices $(p_1,q_1)$ for each DL/UL user pair are illustrated in Fig.~\ref{SAS_Index}.

For the $1\times2$ sub-array configuration, the optimization algorithm predominantly selects the farther sub-arrays: 94\% of DL user~1 and 78\% of UL user~1 selections fall within the sub-array index range [17, 32]. 
Similarly, 92\% of DL user~2 and 71\% of UL user~2 also select sub-arrays within this region. 
Within this range, the SI channel matrix exhibits an average coupling level of $-66.1$ dB, with all Tx-Rx element-level MC values better than $-48.1$ dB. 
Furthermore, 59\% of the selected Tx sub-arrays are concentrated in the [25, 32] range, i.e., the farthest columns from the Rx sub-arrays.

Moreover, for the $1\times4$ sub-array configuration, a similar preference for farther sub-arrays is observed: 96\% of DL user~1 and 88\% of UL user~1 selections are within the index range [9, 16]. Likewise, 94\% of DL user~2 and 63\% of UL user~2 also fall into this range.

\subsection{Sub-array Configuration Impacts on the SI Suppression}
\begin{figure}[!t]
\centerline{\includegraphics[width=\linewidth]{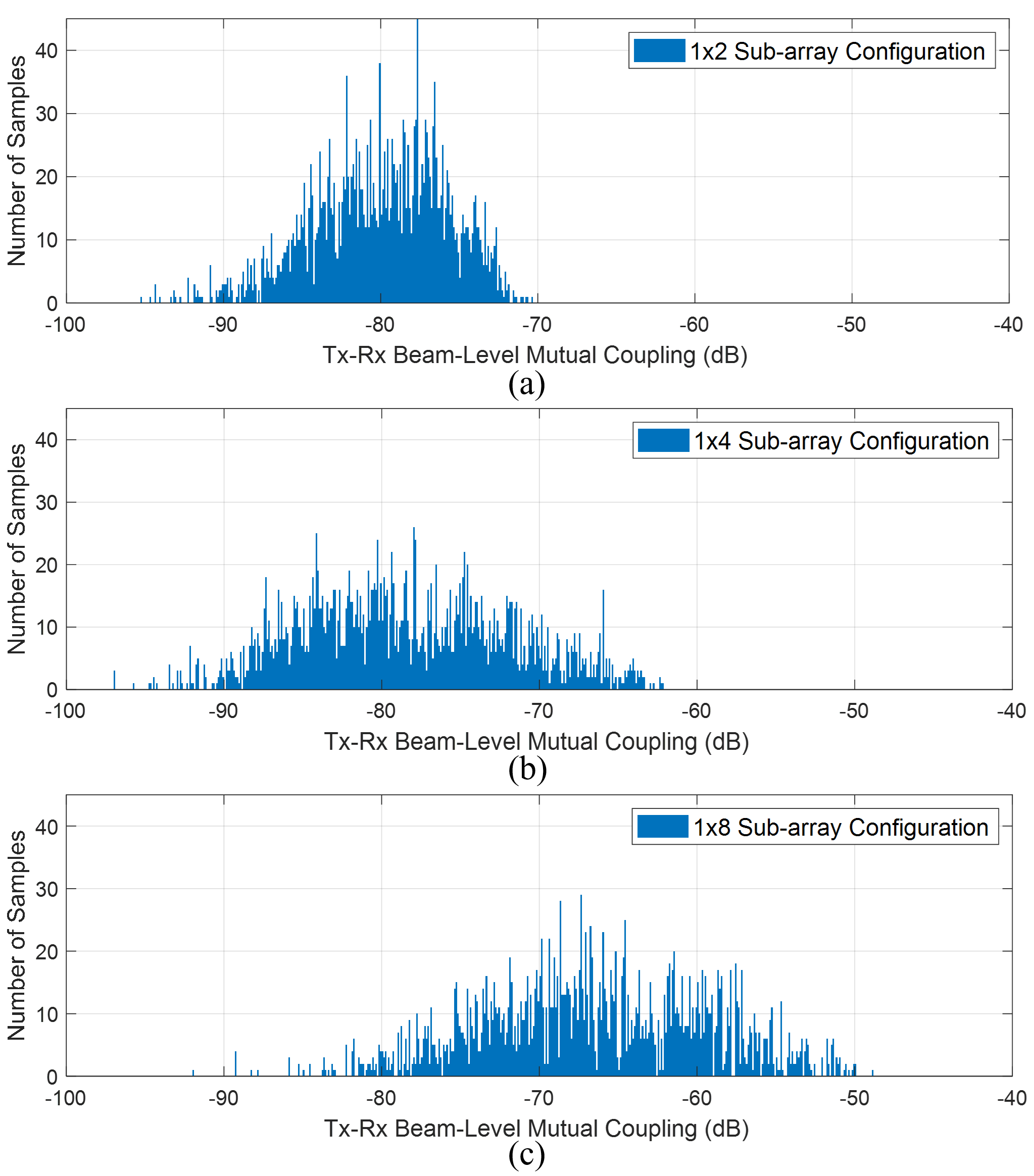}}
\vspace{-0.2cm}
\caption{Tx-Rx Beam-level mutual coupling distribution with proposed angular perturbation-based null-space projection and sub-array selection optimization using sub-array configuration (a) 1$\times$2, (a) 1$\times$4, and (c) 1$\times$8 sub-arrays.}
\label{SI_distribution}
\end{figure}

To comprehensively evaluate the proposed beamforming optimization in terms of SI suppression, both DL/UL user~1 and~2 are assumed to vary within the angular range $\{\psi_{D_1}, \psi_{U_1},\psi_{D_2}, \psi_{U_2}\} \in [0^\circ:30^\circ:180^\circ]$.
The distributions of residual SI after RF beamforming modification with the proposed joint SAS-HBF optimization for 2,401 distinct DL/UL user location combinations are presented in Fig.~\ref{SI_distribution}. 
A comparison between Fig.~\ref{SI_distribution}~(b) ($1\times4$) and Fig.~\ref{SI_distribution}~(c) ($1\times8$) reveals that the $1\times4$ sub-array achieves an $11.9$ dB higher average beam-level isolation, primarily due to its ability to select sub-array pairs with greater separation and thus lower-power SI channels. 
The median isolation likewise improves by $12.1$ dB (from $-66.9$ dB with $1\times8$ to $-79.0$ dB), while the worst-case MC is reduced (improved) by $13.2$ dB, ensuring that the beam-level SI across all DL/UL location combinations remains below $-62.1$ dB.

Further reducing the sub-array length to $1\times2$ introduces even greater flexibility in selecting low-SI channels, resulting in a tightly clustered distribution around a median of $-79.6$ dB, as shown in Fig.~\ref{SI_distribution}~(a). Although its average and best-case isolation are comparable to those of the $1\times4$ configuration, the worst-case MC improves by $8.2$ dB relative to $1\times4$ and by $21.4$ dB relative to $1\times8$, guaranteeing beam-level isolation better than $-70.3$ dB across all sampled DL/UL locations. This narrow spread underscores how increased sub-array selection freedom not only enhances uniformity but also strengthens worst-case SI suppression across diverse user geometries.
The best beam-level isolation achieved across the three tested array configurations is comparable, with differences of less than $5$ dB. The best RF isolation values are $-95.2$ dB, $-96.9$ dB, and $-91.9$ dB for the $1\times2$, $1\times4$, and $1\times8$ sub-array configurations, respectively.

In summary, two clear trends emerge:
\begin{itemize}
\item Sub-array choice among channels with inherently lower SI yields a pronounced shift toward higher average and median beam-level isolation.
\item Increased selection flexibility among comparably performing channels produces more uniform SI suppression performance and markedly improves the worst-case isolation.
\end{itemize}

Additionally, as presented in the previous Section \ref{sub_indices}, sub-arrays with larger spatial separation and a greater number of isolation-enhancement structures are preferable in the selection process to optimize SI suppression performance in FD communications.
However, the improvement in beam-level isolation is not solely attributed to the availability of more distant sub-arrays with inherently stronger SI channel suppression. When comparing the proposed joint SAS-HBF optimization scheme under $1\times4$ and $1\times8$ sub-array configurations, an average gain of $11.9$ dB is achieved. Among this, only $6.6$ dB arises from the increased spatial separation between the farthest columns of the $1\times4$ sub-arrays (i.e., Tx elements 33 to 64 and Rx elements 33 to 64). The remaining improvement stems from the enhanced SI channel selectivity that provides greater design flexibility in the HBF optimization.

Using fewer active elements reduces the effective aperture and coherent beamforming gain. For example, reducing the active aperture from \(1\times8\) to \(1\times4\) and \(1\times2\) results in an approximate maximum array-gain loss of 3 dB and 6 dB, respectively, and may also reduce the coverage advantage of a larger array.
However, FD systems are fundamentally SI-limited, and coverage is governed more by the effective SINR than by beamforming gain alone. In this regime, the illustrative results show that the SI-suppression gain achieved by SAS outweighs the loss in desired-link beamforming gain, so that smaller sub-arrays still provide a net improvement in received SINR and, consequently, overall FD system performance. More importantly, SAS compensates for the reduced aperture by exploiting the diversity of the SI sub-channels, thereby substantially enhancing the achievable SI suppression. Among the three sub-array configurations, the \(1\times4\) sub-array provides a favourable compromise by preserving sufficient array gain and coverage while still offering strong selection diversity and pronounced SI mitigation.

\subsection{SI Channel Correlation and Magnitude Distribution}
\begin{figure}[!t]
\centerline{\includegraphics[width=\linewidth]{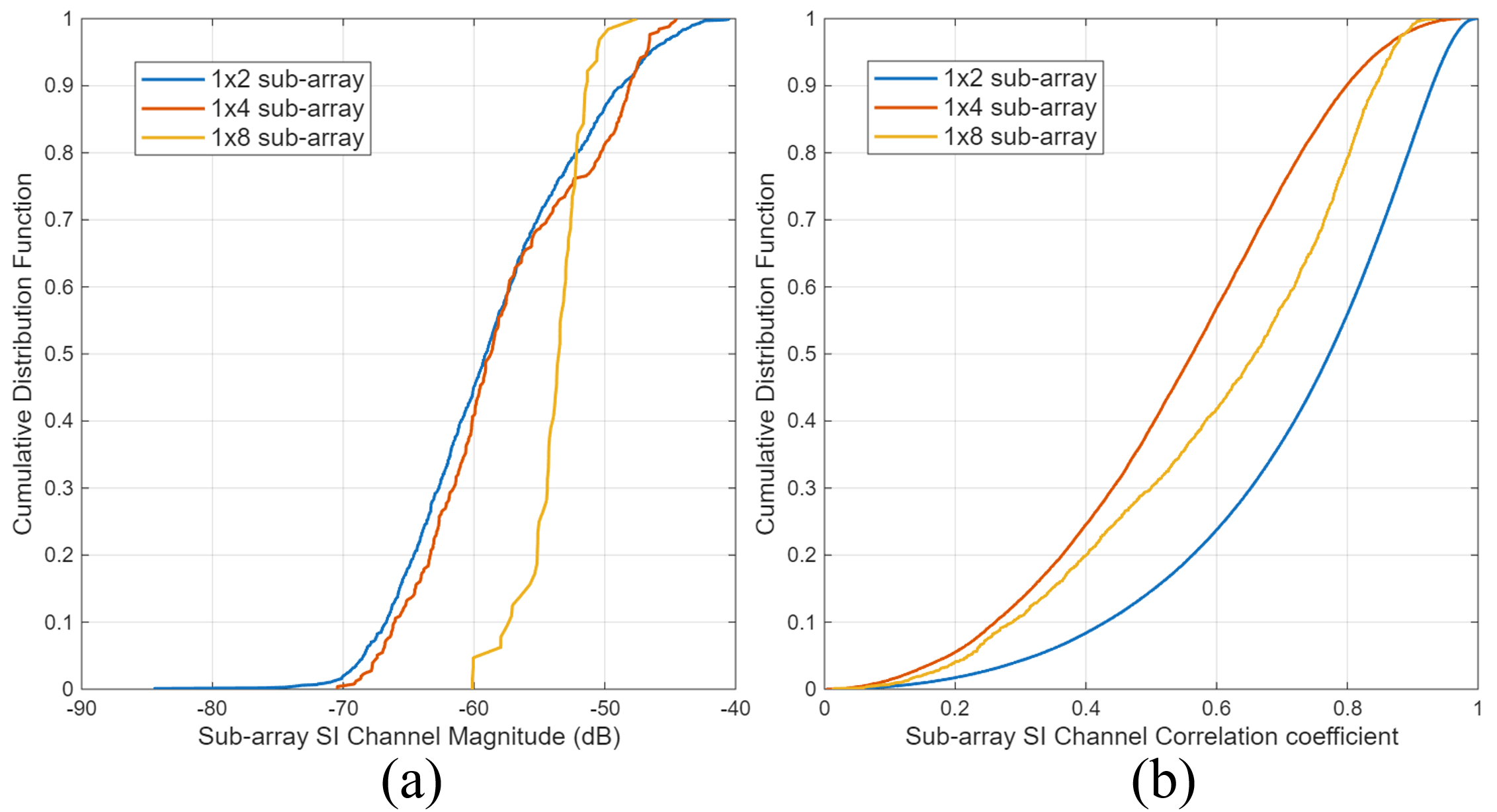}}
\vspace{-0.2cm}
\caption{Various sub-array configurations (a) SI magnitude distribution, (b) SI sub-channel correlation distribution.}
\label{SI_correlation}
\end{figure}
The effectiveness of SAS for SI mitigation in FD communications is mainly determined by the joint effects of: i) the distribution of the SI-channel magnitudes across the candidate sub-arrays, and ii) the spatial correlation among the candidate SI sub-channels. In general, SAS becomes effective when different Tx and Rx sub-array pairs experience sufficiently distinct SI-coupling conditions. In particular, when the candidate sub-channels are weakly or moderately correlated and exhibit noticeable variation in SI magnitude, both the SI channel and the resulting residual SI vary significantly across the candidate pairs, thereby enabling the optimizer to identify sub-array combinations with substantially lower effective SI.

The SI-channel magnitude variations and the correlation distributions for the $1\times2$, $1\times4$, and $1\times8$ sub-array configurations are shown in Fig.~\ref{SI_correlation}. As shown in Fig.~\ref{SI_correlation}~(a), the shorter sub-arrays, such as the $1\times2$ and $1\times4$ configurations, offer greater flexibility in selecting antenna elements with larger spatial separations, and therefore exhibit SI-magnitude distributions that are significantly left-shifted and lower relative to that of the $1\times8$ sub-array. Moreover, the $1\times2$ and $1\times4$ cases exhibit a broader spread and a more pronounced low-SI tail, implying that a large number of low-coupling Tx-Rx sub-array pairs are available for selection. By contrast, the $1\times8$ case exhibits a much steeper cumulative distribution function (CDF) concentrated over a narrower range, closer to the average SI-channel level across all Tx-Rx element pairs, indicating that most candidate sub-array pairs experience similar SI magnitudes and therefore provide less selection gain.

Moreover, Fig.~\ref{SI_correlation}~(b) demonstrates that the SI sub-channels associated with different sub-array pairs exhibit markedly different correlation characteristics. In particular, the $1\times4$ case is the most left-shifted, indicating the lowest correlation among candidate SI sub-channels and hence the strongest selection diversity. This suggests that the $1\times4$ configuration achieves a favourable trade-off between SI-channel magnitude variation and spatial diversity. When the sub-array length is further reduced to the $1\times2$ case, the correlation performance degrades, mainly due to its short sub-array aperture. Nevertheless, its SAS effectiveness for SI suppression still benefits from the broad SI-magnitude distribution and the large number of low-SI candidate pairs, which together preserve strong selection capability. The $1\times8$ case lies between these two in terms of correlation; however, when combined with its much narrower SI-magnitude distribution, it still offers less selection diversity and lower effectiveness than the smaller sub-array configurations.

\subsection{Sub-array Configuration Impacts on the MUI Suppression Performance}
\begin{figure*}[!t]
\centerline{\includegraphics[width=\textwidth]{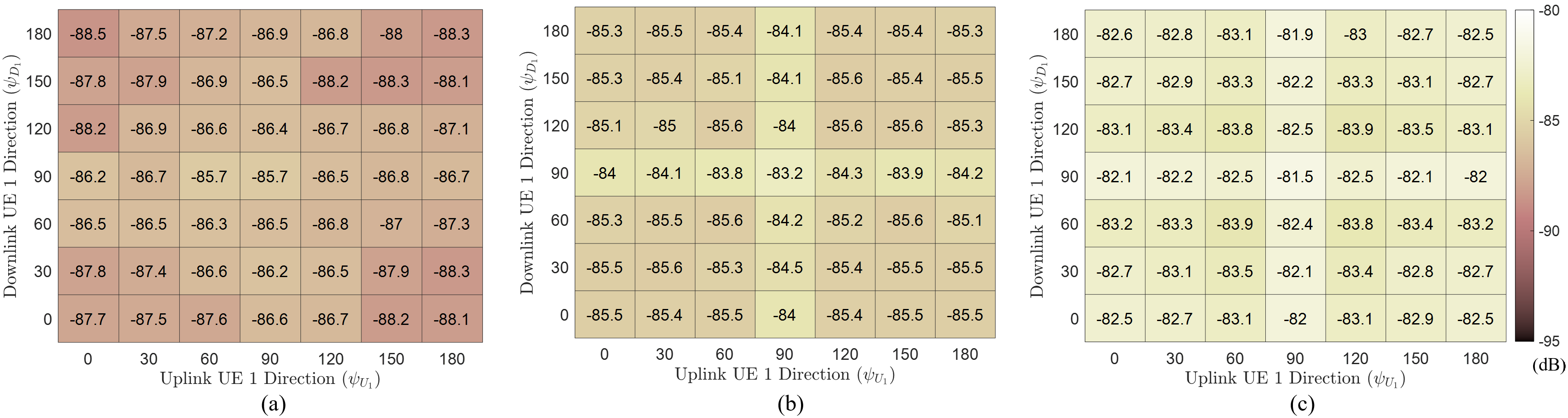}}
\caption{Achievable MUI level (dB) with proposed joint sub-array selection and angular perturbation-based null-space projection beamforming optimization with sub-array configuration (a) 1$\times$2, (a) 1$\times$4, and (c) 1$\times$8 sub-arrays.}
\label{MUI_no_BB}
\end{figure*}

The change in array configuration alters the radiation patterns and beam resolution, thereby affecting the MUI performance. 
To assess this impact, Fig.~\ref{MUI_no_BB} presents the achievable MUI distributions obtained with the proposed joint SAS-HBF optimization scheme across 49 distinct DL/UL user-location combinations for three sub-array configurations: $1\times2$, $1\times4$, and $1\times8$. 
Thanks to the effective beam perturbation and null-projection optimization, the MUI suppression performance exhibits only minor dependence on user locations.
The distribution across DL/UL user combinations remains relatively uniform, with performance variations of only $2.8$ dB for the $1\times2$ configuration and $2.4$ dB for the $1\times8$ configuration.
Across all tested sub-array configurations, beam-level MUI remains below $-81.5$ dB. 

Specifically, without the baseband stage, the $1\times2$ configuration achieves MUI levels better than $-85.7$ dB, with an average of $-87.1$ dB, a median of $-86.9$ dB, and a best-case of $-88.5$ dB. 
As the array length increases, the associated signal gain shift leads to higher interference levels. For the $1\times4$ configuration, the average and median MUI levels degrade to $-85.0$ dB and $-85.4$ dB, respectively, corresponding to an average 2.1 dB increase compared with $1\times2$. Despite this degradation, the worst-case interference remains below $-83.2$ dB across all user combinations.
Further extending the array length to $1\times8$ results in an additional $2.2$ dB increase in beam-level MUI, yielding both an average and median of $-82.8$ dB, with the worst-case interference rising to $-81.5$ dB.

At the same time, when considering the MUI normalized by the desired signal power, where directivity improves with increasing array length, an achievable MUI difference of less than $1.7$ dB is observed across the three configurations, indicating consistent suppression performance of the proposed beam perturbation and null-space projection algorithm regardless of array size.

\subsection{Residual MUI with Baseband Precoder}
\begin{figure*}[!t]
\centerline{\includegraphics[width=\textwidth]{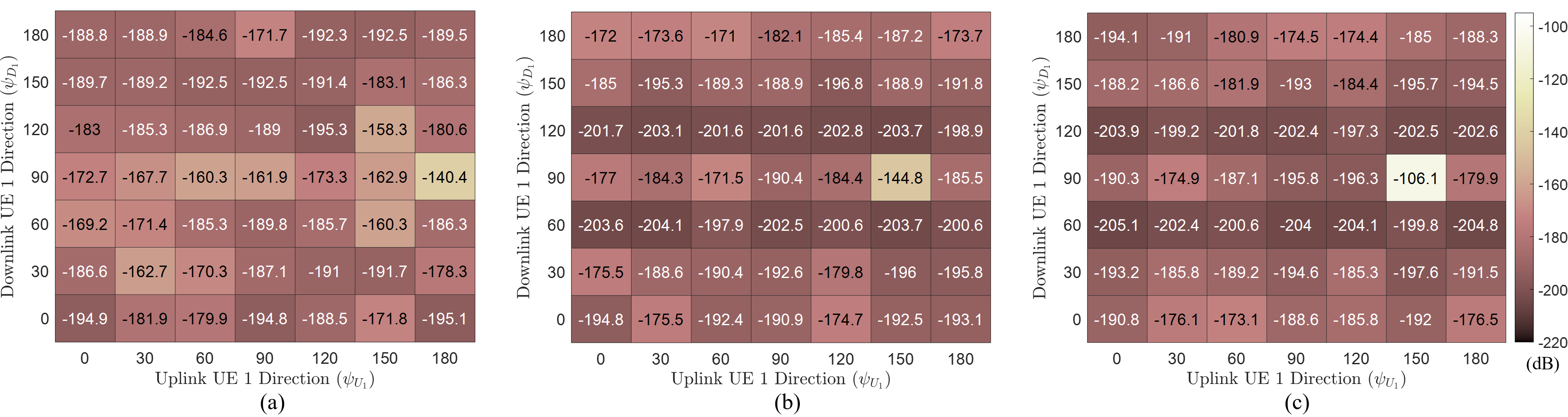}}
\vspace{-0.2cm}
\caption{Achievable MUI level (dB) with proposed joint sub-array selection and angular perturbation-based null-space projection beamforming optimization, and RZF baseband precoder with sub-array configuration (a) 1$\times$2, (b) 1$\times$4, and (c) 1$\times$8 sub-arrays. }
\label{MUI_BB}
\end{figure*}

Fig.~\ref{MUI_BB} presents the residual MUI power achieved by the proposed joint SAS-HBF scheme with RZF baseband precoder across 49 DL/UL user-location combinations. 
Compared with the interference level just after the RF-stage, the baseband-stage RZF precoder provides substantial additional suppression across all sub-array configurations. 
In particular, the $1\times4$ and $1\times8$ cases achieve nearly identical performance, with average MUI levels of $-189.6$ dB. 
By contrast, the $1\times2$ configuration exhibits slightly weaker suppression performance, with an average MUI level of $-181.3$ dB, corresponding to an $8.3$ dB degradation relative to the two larger sub-arrays. 
Across the three schemes, all configurations achieve a median MUI level better than $-185.7$ dB, and except for one DL/UL user-location instance in the $1\times8$ case, all achieve suppression better than $-140.4$ dB. Thanks to the effective MUI suppression achieved by the angular perturbation and null-space projection beamformer, together with the RZF baseband precoder, MUI can be successfully brought to the noise floor and does not pose a bottleneck in FD mMIMO.

\section{Conclusion}\label{Conclusions}
This work has introduced a joint antenna SAS and HBF optimization framework for FD mMIMO systems, targeting simultaneous suppression of SI and MUI. Through an extensive OTA SI channel measurement campaign with an $8\times8$ Tx-$8\times8$ Rx FD array prototype, the study revealed strong spatial variability and reconfigurability of SI channels across different sub-array configurations. By leveraging these variations, the proposed PSO-based joint optimization of sub-array indices and perturbed steering angles demonstrated significant improvements in beam-level isolation.
The results confirm that selecting sub-arrays with inherently lower SI channels substantially enhances isolation performance, while maintaining robustness across diverse DL/UL locations. Illustrative results with measured SI channels showed residual SI suppression improvements of $29.2$ dB and $26.6$ dB for the $1\times2$ and $1\times4$ sub-arrays, respectively, with worst-case gains exceeding $30.7$ dB. Moreover, the joint SAS-HBF scheme achieved average beam-level isolation of $85.2$ dB and $83.3$ dB, while the incorporation of a baseband precoder ensured MUI suppression better than $-181.3$ dB across all configurations.
Working with baseband SIC techniques, the proposed scheme can effectively reduce both SI and MUI close to the noise floor, thereby eliminating key interference bottlenecks and enabling reliable FD mMIMO operation. These findings highlight the potential of joint SAS and HBF optimization as a practical and scalable solution for future FD mMIMO systems.

\bibliographystyle{IEEEtran}
\bibliography{bibYG_2412} 

\begin{IEEEbiography}[{\includegraphics[width=1in,height=1.25in,clip,keepaspectratio]{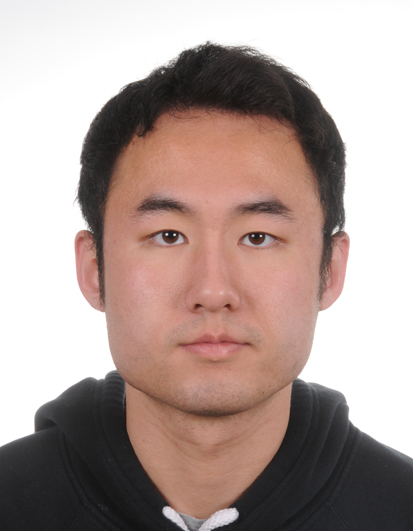}}] {Yuanzhe Gong}
(Member, IEEE) received the B.Eng. (Hons.) and Ph.D. degrees in Electrical and Computer Engineering from McGill University, Montreal, QC, Canada, in 2020 and 2025, respectively. Since 2018, he has served as a Teaching Assistant in the Department of Electrical and Computer Engineering at McGill and as a Research Associate with the Broadband Communication Research Laboratory. He was the recipient of the McGill Engineering Doctoral Award (MEDA), the Graduate Research Enhancement and Travel Award (GREAT Award), the McGill Graduate Excellence Fellowship, the Dean’s Honour List, and the McGill Faculty of Engineering Scholarship. His research interests include wireless communications, antenna design, metamaterial-based large-scale array structures, massive MIMO, near-field communications, intelligent beamforming optimization algorithms, and full-duplex systems.
\end{IEEEbiography}

\begin{IEEEbiography}[{\includegraphics[width=1in,height=1.25in,clip,keepaspectratio]{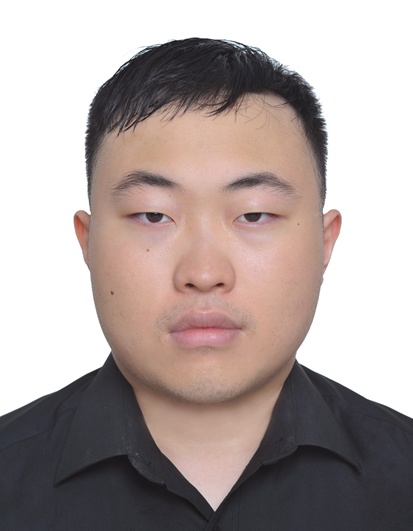}}] {Yuanxing Zhang}
(Graduate Student Member, IEEE) received the B.A.Sc. degree (with Distinction) in Electrical Engineering from the University of British Columbia, Okanagan, Canada, and the M.Sc. (Thesis) degree in Electrical and Computer Engineering from McGill University, Montreal, QC, Canada. He is conducting research with the Broadband Communications Research Laboratory. He was the recipient of the Deputy Vice-Chancellor Scholarship for International Students, and was named to the Dean’s Honour List at UBCO. At McGill, he received the Graduate Research Enhancement and Travel Award (GREAT Award). His research interests include full-duplex massive MIMO systems, hybrid beamforming, sub-array selection, self- and multi-user interference suppression, and beamforming perturbation analysis.
\end{IEEEbiography}

\begin{IEEEbiography}[{\includegraphics[width=1in,height=1.25in,clip,keepaspectratio]{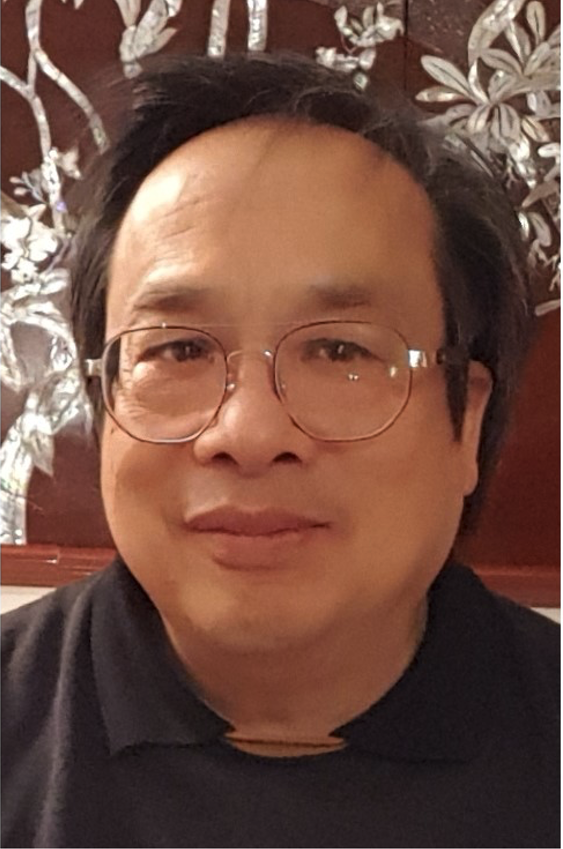}}]{Tho Le-Ngoc} (Life Fellow, IEEE) received the B.Eng. degree in electrical engineering, in 1976, the M.Eng. degree in microprocessor applications, in 1978, from McGill University, Montreal, and the Ph.D. degree in digital communications, in 1983, from the University of Ottawa, Canada. From 1977 to 1982, he was with Spar Aerospace Ltd., Sainte-Anne-de-Bellevue, QC, Canada, involved in the development and design of satellite communications systems. From 1982 to 1985, he was with SRTelecom Inc., Saint-Laurent, QC, Canada, where he developed the new point-to-multipoint DA-TDMA/TDM Subscriber Radio System SR500. From 1985 to 2000, he was a Professor with the Department of Electrical and Computer Engineering, Concordia University, Montreal. Since 2000, he has been with the Department of Electrical and Computer Engineering, McGill University. His research interest includes broadband digital communications.
He is a Distinguished James McGill Professor, and a Fellow of the Engineering Institute of Canada, the Canadian Academy of Engineering, and the Royal Society of Canada. He was a recipient of the 2004 Canadian Award in Telecommunications Research and the IEEE Canada Fessenden Award, in 2005.
\end{IEEEbiography}

\vfill

\end{document}